\documentstyle[preprint,aps,harvard,floats]{revtex}
\newcommand{\half}{\mbox{$\textstyle \frac{1}{2}$}}
\newcommand{\ket}[1]{\left | \, #1 \right \rangle}
\newcommand{\bra}[1]{\left \langle #1 \, \right |}

\newcommand{\beq}{\begin{equation}}
\newcommand{\eeq}{\end{equation}}
\citationstyle{dcu}
\begin{document}
\tightenlines

\title{Quantum Tunneling in Nuclear Fusion}
\author{A.B. Balantekin}
\address{Department of Physics, University of Wisconsin-Madison\\
         Madison, WI 53706 U.S.A\\}
\author{N. Takigawa}
\address{Department of Physics, Tohoku University\\
                980-77 Sendai  Japan\\}
\maketitle
\begin{abstract}
Recent theoretical advances in the study of heavy ion
fusion reactions below the Coulomb barrier are reviewed. Particular
emphasis is given to new ways of analyzing data, such as studying
barrier distributions; new approaches to channel coupling, such as the
path integral and Green function formalisms; and alternative methods
to describe nuclear structure effects, such as those using the
Interacting Boson Model. The roles of nucleon transfer, asymmetry
effects, higher-order couplings, and shape-phase transitions are
elucidated. The current status of the fusion of unstable nuclei and
very massive systems are briefly discussed.
\end{abstract}
\newpage
\tableofcontents
\newpage
\section{Introduction}
\label{I} 
Quantum tunneling in systems with many degrees of freedom is one of
the fundamental problems in physics and chemistry
\cite{hanggi,jjap}. One example of tunneling phenomena in nuclear
physics is the fusion of two nuclei at very low energies. These
reactions are not only of central importance for stellar energy
production and nucleosynthesis, but they also provide new insights
into reaction dynamics and nuclear structure. Until about fifteen
years ago, low energy fusion reactions were analyzed in terms of a
simple model, where one starts with a local, one-dimensional real
potential barrier formed by the attractive nuclear and repulsive
Coulomb interactions and assumes that the absorption into the fusion
channel takes place at the region inside the barrier after the quantum
tunneling.  The shape, location, and the height of this potential were
described in terms of few parameters which were varied to fit the
measured cross sections. The systematics of potentials obtained in
this way were discussed by~\citeasnoun{VAS}. A number of experiments
performed in the early 80's showed that the subbarrier fusion cross
sections for intermediate-mass systems are much larger than those
expected from such a simple picture \cite{beck1,vand1}.  The
inadequacy of the one-dimensional model for subbarrier fusion was
explicitly demonstrated by inverting the experimental data to directly
obtain the effective one-dimensional fusion barrier under the
constraint that it is energy independent \cite{bkn}.
 
In recent years there has been a lot of experimental effort in
measuring fusion cross sections and moments of compound nucleus
angular momentum distributions. A complete compilation of the recent
data is beyond the scope of this review. Several recent reviews
\cite{stef94,reisdorfrev} present an excellent overview of the current
experimental situation. This article reviews theoretical developments
of the last decade in our understanding the multidimensional quantum
tunneling nature of subbarrier fusion.  We should emphasize that the
large volume of new data and extensive theoretical work does not allow
us to provide an exhaustive set of references. The selection we made
does not imply that omitted references are any less important than the
ones we chose to highlight in discussing different aspects of fusion
phenomena.

The natural language to study fusion reactions below the Coulomb
barrier is the coupled-channels formalism. In the last decade 
coupled channels analysis
of the data and the realization of the connection between energy
derivatives of the cross section and the barrier distributions
\cite{neil} motivated accumulation of very high precision data.
 
When the enhancement of the cross section below the barrier was first
observed, many authors pointed out that it is not easy to identify the
underlying physical mechanism \cite{brink,krappe}. Any coupling
introduced between translational motion and internal degrees of
freedom enhances the cross section. The recent high-precision data
helped resolve some of these ambiguities by studying barrier
distributions and mean angular momenta as well. The quality of the
existing data now makes it possible to quantitatively explore a number
of theoretical issues in the quantum tunneling aspects of subbarrier
fusion, such as effects of anharmonic and non-linear interaction
terms.

In the next section, after briefly reviewing observables accessible in
heavy-ion fusion reactions, and the reasons why a one-dimensional
description fails, we present the motivation for studying barrier
distributions. In Section \ref{III} we discuss the standard
coupled-channels formalism, and alternative approaches such as the
path integral formalism and Green's function approaches along with
their various limiting cases. Section \ref{IV} has a dual purpose: it
covers recent attempts to describe nuclear structure effects using the
Interacting Boson Model while illustrating the technical details of
these alternative approaches discussed in Section \ref{III}. In
Section \ref{V}, we survey a representative sample of recent
high-quality data with a focus on new physics insights. We conclude in
Section \ref{VI} with a brief discussion of the open problems and
outlook for the near future.

\section{One-Dimensional Model for Fusion}
\label{II}

\subsection{Experimental Observables}
\label{IIA}

In the study of fusion reactions below the Coulomb barrier the 
experimental observables are the cross section
\begin{equation}
\sigma (E) = \sum_{\ell=0}^{\infty} \sigma_{\ell} (E),
\label{eq1}
 \end{equation}
and the average angular momenta
\begin{equation}
\left\langle \ell (E) \right\rangle = {\sum_{\ell=0}^{\infty} \ell 
\sigma_{\ell} (E) \over  \sum_{\ell=0}^{\infty} \sigma_{\ell} (E)}. 
\label{eq2}
\end{equation}
The partial-wave cross sections are given by 
\begin{equation}
\sigma_{\ell} (E) = {\pi \hbar^2 \over 2 \mu E} (2 \ell +1) T_{\ell} (E), 
\label{eq3}
\end{equation}
where $T_{\ell} (E)$ is the quantum-mechanical transmission
probability through the potential barrier and $\mu$ is the reduced 
mass of the projectile and target system.

Fusion cross sections at low energies are measured by detecting
evaporation residues or fission products from compound nucleus
formation. Evaporation residues can be detected directly by measuring
the difference in velocities between them and the beam-like
ions. Velocity selection can be achieved either by electrostatic
deflectors or by velocity filters. Alternative techniques include
detecting either direct or delayed X-rays and gamma rays. A review of
different experimental techniques to measure the fusion cross sections
is given by \citeasnoun{beck1}.

Several techniques were developed for measuring moments of the angular
momentum distributions. The advent of detector arrays enabled
measurement of full gamma-ray multiplicities \cite{fisch,mel}, 
which, under some assumptions, 
can be converted to $\sigma_{\ell}$ distributions. It is also possible
to measure relative populations of the ground and isomeric states in
the evaporation residues to deduce the spin distribution in the
compound nucleus \cite{stok3,diG90}. Finally, the anisotropy of the
fission fragment angular distribution can be related to the second
moment of the spin distribution \cite{back85,vand2,vand3}. 

It is worthwhile to emphasize that moments of angular momenta, unlike 
the fusion cross section itself, are not directly measurable 
quantities. One needs to make a number of assumptions to convert 
gamma-ray multiplicities or isomeric state populations to average 
angular momenta. An elaboration of these assumptions along with 
a thorough discussion of the experimental techniques is given by
\citeasnoun{vand1}.

\subsection{One-Dimensional Barrier Penetration}
\label{IIAB}

The total potential between the target and projectile nuclei 
for the $\ell$-th partial wave is given by
\begin{equation}
\label{vsubl}
V_\ell(r) = V_N(r) + V_C(r) + \frac{\hbar^2 \ell(\ell+1)}{2\mu r^2} 
= V_0(r) + \frac{\hbar^2 \ell(\ell+1)}{2\mu r^2} \,,
\end{equation}
where $V_N$ and $V_C$ are the nuclear and Coulomb potentials,
respectively. The $\ell =0$ barrier is referred to as the
Coulomb barrier. The barriers obtained in Eq. (\ref{vsubl}) is 
illustrated in Figure \ref{Fig0} for several $\ell$-values. 

For a one-dimensional barrier transmission probabilities
can be evaluated numerically - either exactly or using a uniform WKB
approximation, valid for energies both above and below the barrier
\cite{brink1,brnk} :
\begin{equation}
\label{4}
T_{\ell}(E) = \left[ 1 + \exp \left( 2S_{\ell}(E) \right)\right]^{-1} ,
\end{equation}
where the WKB penetration integral is
\begin{equation}
S_\ell(E) = \sqrt{2\mu \over \hbar^2} \int_{r_{1\ell}}^{r_{2\ell}} dr
\left[ V_0(r) + {\hbar^2 \ell(\ell+1) \over 2\mu r^2} - E \right]^{1/2} .
\label{5}
\end{equation}
In this equation $r_{1\ell}$ and $r_{2\ell}$ are the classical turning
points for the $\ell$-th partial wave potential barrier.

If we assume that the potential barrier can be replaced by a parabola
\beq
V_0(r)=V_{B0} - \half \mu^2 \Omega^2 (r-r_o)^2,
\label{parabola}
\eeq 
where $V_{B0}$ is the height and $\Omega$ is a measure of the
curvature of the s-wave potential barrier, 
the transmission probability can be calculated to be 
\cite{hilwhe}
\begin{equation}
  \label{eq:hillw}
  T_0(E) =  \left[ 1 + \exp
\left[- {2 \pi \over \hbar\Omega} (E-V_{B0}) \right] \right]^{-1}.
\end{equation}
In the nuclear physics literature Eq. (\ref{eq:hillw}) is known as the
Hill-Wheeler formula. Especially at energies well below the barrier
there are significant deviations from this formula as the parabolic
approximation no longer holds.

\subsection{Barrier Distributions}
\label{IIB}

An alternative way of plotting the total cross section data is to look
at the second energy derivative of the quantity $E\sigma$, sometimes
called the distribution of the barriers.  To elaborate on the physical
significance of this quantity let us consider penetration
probabilities for different partial waves in the case of a
one-dimensional system (coupling to an internal system is
neglected), Eq. (\ref{4}).  Under certain
conditions, to be elaborated in the next subsection, we can
approximate the $\ell$ dependence of the transmission probability at a
given energy by simply shifting the energy \cite{bkn,reimer}:
\begin{equation}
T_{\ell} \simeq T_0 \left[ E - \frac{\ell (\ell +1) \hbar^2}{2 \mu
R^2(E)} \right],
\label{6}
\end{equation}
where $\mu R^2(E)$ characterizes an effective moment of inertia. 
$R(E)$ was found to be a slowly varying function of
energy as depicted in Figure \ref{Fig1}. Consequently, in many
applications $R(E)$ is replaced by $r_0$, the position of the s-wave
barrier, in Eq. (\ref{6}).  If many values of $\ell$ are important in
the sum over partial-wave transmission probabilities in Eq. (1), we
can approximate that sum with an integral over $\ell$, and, using
Eq. (\ref{6}) obtain \cite{bkn,reimer}
\begin{equation}
E \sigma(E) = \pi R^2(E) \int_{-\infty}^E dE' T_0(E').
\label{7}
\end{equation}

It was found that Eq. (\ref{7}) represents the experimental data for
the total fusion cross section rather well \cite{reimer,baba,serdar}.
Differentiating Eq. (\ref{7}) twice \cite{neil} one finds that the
energy derivative of the s-wave transmission probability is
approximately proportional to the second energy derivative of the
quantity $E\sigma$ up to corrections coming from the energy dependence
of $R(E)$:
\begin{equation}
{dT_0 (E) \over dE} \sim \frac{1}{\pi R^2(E)} 
\frac{d^2}{dE^2} (E \sigma (E)) + {\cal O}(\frac{dR}{dE}).
\label{8}
\end{equation}
Since $R(E)$ is a slowly varying function of energy, the first term in
Eq. (\ref{8}) can be used to approximate the first derivative of the
s-wave tunneling probability. For a completely classical system, $T_0$
is unity above the barrier and zero below; hence the quantity $dT_0
(E)/dE$ will be a delta function peaked when $E$ is equal to the
barrier height, as shown in Figure \ref{Fig2a}.  Quantum mechanically
this sharp peak is broadened as the transmission probability smoothly
changes from zero at energies far below the barrier to unity at
energies far above the barrier (see Figure
\ref{Fig2a}). \citeasnoun{neil} suggested that if many channels are
coupled to the translational motion, the quantity $dT_0 (E)/dE$ is
further broadened and can be taken to represent the ``distributions of
the barriers'' due to the coupling to the extra degrees of freedom as
depicted in Figure \ref{Fig2b} for a two-channel case. 

Obviously one needs rather high precision data to be able to
numerically calculate the second derivative of the excitation
function. Such high-quality data recently became available for a
number of systems. As a specific example of the quality of recent
data, Figure \ref{Fig3} shows the measured cross section and the
extracted associated barrier distribution for the $^{16}$O $+^{154}$Sm
system \cite{wei,lei95}.

Eqs. (\ref{6}) and (\ref{7}) 
(with the substitution $R(E) \rightarrow r_0$) can
also be used to obtain a direct connection between the fusion cross
section and the angular momentum distribution \cite{sahm,ack95} : 
\beq
T_{\ell}(E) = {1 \over \pi r_0^2} \left( {d(E' \sigma (E')) \over dE'}
\right), \eeq where \beq E' = E - (\hbar^2 / 2\mu r_0^2) \ell(\ell+1).
\eeq

For energies well above the barrier, one can use 
the parabolic approximation of Eq. (\ref{parabola}) for the potential 
barrier.  Further approximating $R(E)$ by
$r_0$ and inserting the penetration probability for the parabolic
barrier, Eq. (\ref{eq:hillw}), in Eq. (\ref{7}) one obtains an
approximate expression for the cross section \cite{wong} 
\beq 
\sigma (E) = {\hbar \Omega r_0^2 \over 2E} \log \left[ 1 + \exp
\left[ {2 \pi \over \hbar\Omega} (E-V_{B0}) \right] \right].
\label{wongformula}
\eeq 
In the classical limit, where $\Omega \rightarrow 0$ or $E \gg
V_{B0}$, Eq. (\ref{wongformula}) reduces to the standard geometrical
result 
\beq 
\sigma (E) = \pi r_0^2 \left( 1 - { V_{B0} \over E} \right).
\label{geometry}
\eeq

\subsection{Energy Dependence of the Effective Radius}
\label{IIc}

If one sets $R(E)=r_0$ in Eq. (\ref{6}) for approximating the 
$\ell$-wave penetrability by the $s$-wave penetrability at a 
shifted energy, one gets only the leading term in what 
is actually an infinite series
expansion in $\Lambda = \ell(\ell+1)$. The next term in this expansion
can easily be calculated.
  Let $r_\ell$ denote the position of the peak of the
$\ell$-wave barrier which satisfies
\begin{equation}
\label{rldef}
\left. {\partial V_\ell(r) \over \partial r} \right|_{r=r_\ell} =0 \,,
\end{equation}
and
\begin{equation}
\left. {\partial^2 V_\ell(r) \over \partial r^2} \right|_{r=r_\ell} <0 \,,
\end{equation}
then the height of the barrier is given by $V_{Bl} = V_l(r_\ell)$.
We make the ansatz that the barrier position can be written as an infinite
series,
\begin{equation}
\label{rlans}
r_\ell = r_0 +  c_1 \Lambda + c_2 \Lambda^2 + \cdots,
\end{equation}
where the $c_i$ are constants.  Expanding all functions in
Eq.~(\ref{rldef}) consistently in powers of $\Lambda$, we find that
the first coefficient is
\begin{equation}
\label{c1}
c_1 = - \, {\hbar^2 \over \mu\alpha r_0^3} \,,
\end{equation}
where $\alpha$ is the curvature of the $s$-wave barrier
\begin{equation}
\alpha= \left. - \, {\partial^2 V_0(r) \over \partial r^2 }
\right|_{r=r_0} \,.
\end{equation}
Substituting the leading order correction in the barrier position
$r_\ell$ into Eq.~(\ref{vsubl}), we find that to second order in
$\Lambda$ the $\ell$-wave barrier height is given by
\begin{equation}
\label{vbl}
V_{Bl} = V_{B0}
+ \frac{\hbar^2 \Lambda}{2\mu r_0^2}
+ \frac{\hbar^4 \Lambda^2}{2\mu^2 \alpha r_0^6} \,.
\end{equation}
Therefore, an improved approximation for the $\ell$-dependence in the
penetrability is given by
\begin{equation}
\label{tlcor}
T_\ell(E) \approx T_0 \left(E - \frac{\hbar^2 \Lambda}{2\mu r_0^2} -
\frac{\hbar^4 \Lambda^2}{2\mu^2 \alpha r_0^6}\right).
\end{equation}
Comparing Eq.~(21) with Eq.~(6), we find that the
energy-dependent effective radius can be expressed as \cite{serdar}
\begin{equation}
\label{reff}
R^2(E) = r_0^2 \left[ 1 - {4 \over \alpha r_0^2} { \int_0^E dE' \,
T_0(E') (E-E') \over \int_0^E dE' \, T_0(E')} \right].
\end{equation}
This expression is useful in assessing the utility of the second
derivative of the quantity $E\sigma$ as the distribution of barriers
as it gives an estimate of the terms neglected in Eq. (\ref{8}).
If we rewrite Eq. (\ref{8}) including previously neglected terms 
\begin{equation}
{dT_0 (E) \over dE} = \frac{1}{\pi R^2(E)} \frac{d^2}{dE^2} 
\left( E \sigma (E) \right) - \frac{E\sigma (E)}{\pi R^4(E)} 
\frac{d^2}{dE^2} \left( R^2(E)\right) - \frac{2 T_0(E)}{R^2(E)} 
\frac{d}{dE} \left( R^2(E)\right), 
\label{yeni8}
\end{equation}
we see that a strong energy dependence of $R(E)$ would not only 
provide an overall multiplicative factor between the experimental 
observable $d^2(E\sigma)/dE^2$ and the true barrier distribution 
(i.e., $dT_0/dE$), but may also shift the position of various 
peaks in it and change the weights of these peaks. 
Eq. (\ref{reff}) can be used to illustrate that such 
corrections are indeed small \cite{serdar}. This is a useful 
consistency check even when channel coupling effects yield a number 
of eigen-barriers (cf. Section \ref{IIIanew}) as Eq. (\ref{8}) 
still needs to be satisfied for each one-dimensional eigen-barrier 
to be able to interpret experimentally determined $d^2(E\sigma)/dE^2$ 
as the distribution of barriers.

\subsection{Inversion of the Data}
\label{IId}
Using Abelian integrals \cite{cole}, one can show that for energies
below the barrier
\begin{equation}
\label{a2}
\int_E^{V_{B\ell}} dE' \, { S_{\ell}(E') \over \sqrt{E'-E}} = {\pi
\over 2} \sqrt{2\mu \over \hbar^2} \int_{r_{1\ell}}^{r_{2\ell}} dr
\left[ V_0(r) + {\hbar^2 \ell(\ell+1) \over 2\mu r^2} - E \right] ,
\end{equation}
where $V_{B\ell}$ is the height of the $\ell$-wave potential and
$V_0(r)$ is the $s$-wave barrier. The energy derivative of
the left-hand side of 
Eq.~(\ref{a2}) can be integrated by parts to yield,
\begin{equation}
\label{a3}
\int_E^{V_{B\ell}} dE' \, { \partial S_{\ell}(E')/\partial E' \over
\sqrt{E'-E}} = - {\pi \over 2} \sqrt{2\mu \over \hbar^2} \; (r_{2\ell}
- r_{1\ell}) \,,
\end{equation}
which is used to find the barrier thickness \cite{bkn}. Using
Eqs. (\ref{4}) and (\ref{6}), one can relate the WKB penetration
integral to the experimentally measured cross section as
\begin{equation}
S_0(E) = {1 \over 2} \log \left[ \left[ {d \over dE} \left[ {E \sigma
(E) \over \pi R^2(E)} \right] \right]^{-1} - 1 \right].
\label{a4}
\end{equation}
Thus, if $R(E)$ is specified, the thickness of the barrier at a given
energy is completely determined from the experimental data using
Eqs. (\ref{a3}) and (\ref{a4}).

The potentials resulting from the analysis of \citeasnoun{bkn} for six
systems are shown in Figure \ref{Fig4}. For comparison, point Coulomb
and the phenomenologically determined potential of \citeasnoun{kns}
are also shown. It should be emphasized that this inversion method 
assumes the existence of a single potential barrier. The thickness 
functions, $t=r_{20}-r_{10}$, in Figure \ref{Fig4} especially
for the heavier systems are inconsistent with the assumption of a
single-valued one-dimensional local potential, clearly indicating the
need for coupling to other degrees of freedom. This result was
confirmed by the systematic study of \citeasnoun{inui}. 

\section{Multidimensional Quantum Tunneling in Nuclear Physics}
\label{III}
\subsection{Coupled-Channels Formalism}
\label{IIIa}
A standard theoretical approach to study the effect of nuclear
intrinsic degrees of freedom on the fusion cross section is to
numerically solve coupled-channels equations that determine the wave
functions of the relative motion. For example, if one is interested in
the effect of the excitation of the ground state $K=0^+$ rotational
band of the target nucleus, then each channel can be denoted by a set
of indices $(I,\ell)$, where $I$ and $\ell $ are the angular momentum
of the rotational excited states of the target nucleus and that of the
relative motion, respectively.  The coupled-channels equations then
read
\begin{eqnarray}
&&\left[\ {-{\hbar^2\over {2\mu}} {d^2\over {dr^2}} + {\hbar^2\over
{2\mu r^2}} \ell_1(\ell_1+1)+V(r)-E_{I_1}} \right] f_{I_1,\ell_1}(r)
\nonumber\\ &+&F_{\lambda}(r) \sum_{I_2,\ell_2} (-)^{J+\ell_2}
i^{I_2+\ell_2-I_1-\ell_1} \left[{(2\ell_1+1)(2I_1+1)(2I_2+1)\over
{2\lambda+1}}\right]^{1/2} \nonumber \\ &\times&
<I_1I_200|\lambda 0><\ell_1\lambda 00|\ell_20> {\ell_1 I_1 J\brace I_2
\ell_2 \lambda} f_{I_2,\ell_2}(r)=0.
\label{III1} 
\end{eqnarray}
When the coupled-channels formalism is used to study direct reactions 
the optical potential $V(r)$ contains an imaginary part in order to
take into account the effect of intrinsic degrees of freedom other
than the rotational excitation on the scattering.  The multi-polarity
of the intrinsic excitation is $\lambda =2$ if we restrict to
quadrupole deformations. The coupling form factor $F_{\lambda}(r)$
consists of Coulomb and nuclear parts,
\begin{equation}
F_{\lambda}(r)=F_C(r)+F_N(r).
\label{III2}
\end{equation}
If the coupling is restricted to the quadrupole deformations, the
Coulomb part is
\begin{eqnarray}
F_C(r)&=&{3\over\sqrt{20\pi}}\beta\; Z_1Z_2e^2{R_c^2\over
r^3}\qquad\qquad (r\geq R_c) \nonumber \\
&=&{3\over\sqrt{20\pi}}\beta\; Z_1Z_2e^2{r^2\over R_c^3}\qquad\qquad
(r<R_c)
\end{eqnarray}
and the nuclear part is
\begin{equation}
F_N(r)=-{\sqrt{5 \over 4 \pi}}\beta R_V {dV_N \over dr}\quad.
\label{III4}
\end{equation}
In most calculations the scale parameters $R_c$ and $R_V$ are taken to
be $1.2A^{1/3}$. In the standard coupled-channels calculations, one
solves Eq. (\ref{III1}) by imposing regular boundary condition at the
origin. In contrast, in the study of heavy-ion fusion reactions, one 
takes the potential to be real and 
often solves Eq. (\ref{III1}) by imposing the incoming wave boundary
condition at some point inside the potential barrier to obtain
S-matrices.  The fusion cross section is then obtained based on the
unitarity relation as \beq \sigma={\pi\over k^2} \sum_{\ell}(2\ell
+1)\{1-\sum_a \vert S_a(\ell)\vert^2\},
\label{III5}
\eeq 
where the index $a$ designates different scattering channels, which
have been explicitly dealt with in the coupled-channels calculation.
An advantage of the coupled-channels method is that one can try to
consistently analyze heavy-ion fusion reactions, and other scattering
processes such as elastic scattering. Coupled-channels calculations
for a number of systems exist in the literature
\cite{dasso,lindsay,esbensen,thomp,stef1}.

One can significantly reduce the dimension of the coupled-channels
calculations by ignoring the change of the centrifugal potential
barrier due to the finite multi-polarity of the nuclear intrinsic
excitation \cite{nob1,nob3}. This is called the no-Coriolis
approximation, rotating frame approximation, or iso-centrifugal
approximation \cite{gomez}. A path integral approach to no-Coriolis
approximation was given by \citeasnoun{hag95}.

If one further ignores the finite excitation energy of nuclear
intrinsic motion, then one can completely decouple the
coupled-channels equations into a set of single eigen-channel
problems. These two approximations significantly simplify the
numerical calculations and also give a clear physical understanding of
the effect of channel-coupling in terms of the distribution of
potential barriers.

\subsection{Simplified Coupled-Channels Models}
\label{IIIanew}

Under certain assumptions it is possible to significantly simplify 
the coupled-channels equations described in the previous section. 
We take the Hamiltonian to be 
\begin{equation}
H= H_k + V_0(r) + H_{0}(\xi) + H_{\rm int} ({\bf r},\xi)
\label{29}
\end{equation}
with the term $H_k$ representing the kinetic energy 
\begin{equation}
H_k = -\frac{\hbar^2}{2\mu}\nabla^2, 
\end{equation}
where ${\bf r}$ is the relative coordinate of the colliding 
nuclei and ${\xi}$ represents any internal degrees of freedom of
the target or the projectile. In this equation $V_0(r)$ is the bare
potential and the term $H_0 (\xi )$ represents the internal structure
of the target or the projectile nucleus. 
Introducing the eigenstates of $H_{0}(\xi)$
\begin{equation}
H_0 |n\rangle = \epsilon_n |n\rangle, 
\end{equation}
and expanding the radial wave function as 
\begin{equation}
\Psi (r) = \sum_n \chi_n (r) |n\rangle, 
\end{equation}
the time-independent Schroedinger equation is reduced to a set of 
coupled equations for the relative motion wave functions  $\chi_n$, 
\begin{equation}
\left[ - \frac{\hbar^2}{2 \mu} \frac{d^2}{dr^2} + V_{\ell}(r) - E 
\right]  \chi_n(r) = - \sum_m [ \epsilon_n \delta_{nm} + \langle n | 
H_{\rm int} ( r,\xi ) | m \rangle ]  \chi_m(r) . 
\end{equation}
These equations are solved under the incoming wave boundary 
conditions:
\begin{equation}
\chi_n(r) \rightarrow \left\{ \begin{array}{cc} \delta_{n0} \exp
(-ik_nr) + \sqrt{\frac{k}{k_n}} R_n \exp (ik_nr), \>r \rightarrow +
\infty \\ \\ \sqrt{\frac{k}{k_n}} T_n \exp (-ik_nr),
\>\>\>\>\>\>\>\>\>\>\>\>\>\> \>\>\>\>\>\>\>\>\>\>\>\>\>\>\>\>\> r
\rightarrow r_{min}
\end{array} \right\} , 
\label{iwbc}
\end{equation}
where $\hbar^2 k_n^2 / 2 \mu = E - \epsilon_n$. The internal 
degrees of freedom are taken to be initially in their ground state 
labeled by $n=0$ and the associated ground state energy is set to be 
zero, $\epsilon_0=0$. In Eq. (\ref{iwbc}) the reflection and 
transmission coefficients in each channel are
denoted by $R_n$ and $T_n$, respectively.

Several groups \cite{dasso,broglia,jacobs} studied various 
simplifying limits to emphasize salient physical features. Here we  
summarize the approach of \citeasnoun{dasso}. They assumed that the 
coupling interaction factors into a relative part, $F(r)$, and an 
intrinsic part, $G(\xi)$: 
\begin{equation}
M_{nm} \equiv \epsilon_n \delta_{nm} + 
\langle n | H_{\rm int} ( r,\xi ) | m \rangle  = 
\epsilon_n \delta_{nm} + F(r) \langle n | G(\xi ) | m \rangle ,
\label{mequation} 
\end{equation}
and that the form factor $F(r)$ is a constant (taken by 
\citeasnoun{dasso} to be the value 
of $F(r)$ at the barrier position). Under these approximations the 
coupled-channel equations decouple to give 
\begin{equation}
\left[ - \frac{\hbar^2}{2 \mu} \frac{d^2}{dr^2} + V_{\ell}(r) 
+ \lambda_n - E \right]  \left( \sum_m U_{nm} \chi_m(r) \right) = 0 , 
\label{dassonomanga}
\end{equation}
where $U_{nm}$ is the unitary matrix which diagonalizes the coupling 
matrix $M_{nm}$ to give a set of eigenvalues $\lambda_n$. Eq. 
(\ref{dassonomanga}) indicates that the effect of the coupling is to 
replace the original barrier by a set of eigen-barriers $V_{\ell}(r) 
+ \lambda_n$. The transmission probability calculated in the 
incoming-wave boundary conditions is given by 
\begin{equation}
T_{\ell} (E) = \sum_n |U_{n0}|^2 T_{\ell}(E-\lambda_n) , 
\end{equation}
where $T_{\ell}(E-\lambda_n)$ is the transmission probabilities 
calculated at shifted energies $E-\lambda_n$. 

Even though the constant-coupling approach would overpredict the 
transmission probability, it can nevertheless be 
used to get a qualitative 
understanding of the dependence of the fusion cross section on 
various physical quantities. For example, one can study coupling 
to an harmonic mode with a finite Q-value using the model Hamiltonian 
\begin{equation}
M = \pi^2 / 2D + \frac{1}{2} C \xi^2 + F_0 \xi,
\end{equation}
yielding
\begin{equation}
M_{mn}= - n Q \delta_{mn} + F ( \sqrt{n} \delta_{n,m+1} + 
\sqrt{n+1} \delta_{n,m-1}) ,
\end{equation}
where $-Q = \hbar \sqrt{C/D}$ is the excitation energy and $F=F_0 
\sqrt{|Q|/2C}$ measures the total strength of the coupling and the 
eigenvalues are 
\begin{equation}
  \lambda_n = n |Q| - F^2/|Q|.
\end{equation}
The total transmission probability can easily be calculated to be
\begin{equation}
\label{simcc}
  T_{\rm cc}(E) = \sum_{n=0}^{\infty}  ( F^{2n}/ Q^{2n} n!) 
\exp(-F^2/Q^2) T(E-n|Q|+F^2/|Q|). 
\end{equation}
In the special case of a two-channel problem the matrix
\begin{equation}
 M = \left( \begin{array}{cc} 0 & \>\>\>F \\  \\  F & 
\>\>\>-Q \end{array} \right) 
\end{equation}
has eigenvalues
\begin{eqnarray}
  \lambda_{\pm} = \frac{1}{2} \left( -Q \pm \sqrt{Q^2 + 4 F^2} \right),
\end{eqnarray}
with the corresponding weight factors
\begin{equation}
  U^2_{\pm} = \frac{2 F^2}{4F^2 + Q^2 \mp Q \sqrt{4F^2+Q^2}}.
\end{equation}
Note that for $F/|Q| <1$ the lowest effective barrier carries the 
largest weight for negative $Q$, while the situation is reversed for 
positive $Q$ (cf. Section \ref{nuctrans}). 

\citeasnoun{tanimura} pointed out that the constant coupling 
approximation can overestimate low-energy fusion rates where the 
coupling effects are strong and the associated form factors are 
rapidly varying. \citeasnoun{dasso5} extended their model for strong 
coupling cases. They diagonalized the coupled intrinsic system in 
the barrier region to obtain the eigenstates $\ket{\gamma(r)}$ and 
the eigenvalues $\lambda_{\gamma}(r)$ as functions of $r$ to obtain 
the total transmission probability
\begin{equation}
  \label{eq:yalanci}
  T_{\rm scc}(E) = \sum_{\gamma} | \bra{\gamma(R)} 0 \rangle|^2 
T_{\gamma} (E, V(r)+\lambda_{\gamma}(r)),
\end{equation}
where $T_{\gamma} (E, V(r)+\lambda_{\gamma}(r))$ is the penetration 
probability for the potential $ V(r)+\lambda_{\gamma}(r)$. The weighing 
factors are fixed at a chosen radius $R$, which might be the position 
of the unperturbed barrier or the average position of the eigen-barriers. 
Two coupled-channels codes, simplified in
this manner, are available in the literature: CCFUS \cite{dasso2} and
CCDEF \cite{CCDEF}. Both of them treat the vibrational coupling in the
constant coupling approximation.  The latter takes into account
projectile and target deformations within the sudden approximation and
treats coupling to the transfer channels with a constant form factor.

In CCFUS the basis states included are the ground state $\ket{0}$, 
the quadrupole one-phonon state $b_2^{\dagger} \ket{0}$, the 
octupole one-phonon state $b_3^{\dagger} \ket{0}$, and the product 
two-phonon state $b_2^{\dagger} b_3^{\dagger} \ket{0}$. The resulting 
matrix to be diagonalized to yield the eigen-channels is 
\begin{equation}
\label{ccfuspalavra}
M = \left(
  \begin{array}{cccc}
    0 & F_2 (r)& F_3 (r) & 0 \\
    F_2 (r)& \epsilon_2 & 0 & F_3 (r) \\
    F_3 (r) & 0 & \epsilon_3 & F_2 (r)\\
    0  &  F_3 (r) & F_2 (r)& \epsilon_2 + \epsilon_3 
    \end{array} \right) .
\end{equation}
In CCFUS the double phonon states $(b_2^{\dagger})^2 \ket{0}$, and 
$(b_3^{\dagger})^2 \ket{0}$ are not included for mathematical 
simplification. The eigenvalues of the matrix in Eq. 
(\ref{ccfuspalavra}) can be written as the sums of the eigenvalues 
of the two $2\times 2$ matrices which represent the coupling of the 
single one-phonon states, i.e., 
\beq
\label{mat1}
M_2 = \left(
  \begin{array}{cc}
    0 & F_2 (r)\\
    F_2 (r)& \epsilon_2
\end{array} \right) ,
\eeq
and
\beq
\label{mat2}
M_3 = \left(
  \begin{array}{cc}
    0 & F_3 (r) \\
    F_3 (r) & \epsilon_3
\end{array} \right) .
\eeq
If the matrix of Eq. (\ref{ccfuspalavra}) also included the 
double-phonon states $(b_2^{\dagger})^2 \ket{0}$, and 
$(b_3^{\dagger})^2 \ket{0}$ such a simple relationship between 
eigenvalues of the $4\times 4$ and $2\times 2$ matrices 
would not be possible. CCFUS provides two options; 
the matrices of Eq. (\ref{mat1}) and (\ref{mat2}) are either 
diagonalized by replacing the form factors $F_2 (r)$ and 
$F_3 (r)$ with their values at the location of the bare 
potential barrier, or diagonalized for all values of $r$. 
In the latter case only the weight factors, but not the eigenvalues 
are evaluated at the position of the bare barrier. The transmission 
coefficients for each eigen-barrier are calculated using the 
Wong formula, Eq. (\ref{wongformula}). 

The approach of CCFUS can be generalized to incorporate $n$ 
different phonons by including the ground state, $n$ one-phonon 
states, $n(n-1)/2$ two-phonon states (i.e., not those states where 
the same phonon appears more than once), $n(n-1)(n-2)/3!$ 
three-phonon states, etc. to obtain a total number of 
\begin{equation}
  \label{eq:nandasayi}
 1 + n + \frac{n(n-1)}{2!} + \frac{n(n-1)(n-2)}{3!} 
+ \frac{n(n-1)(n-2)(n-3)}{4!} + \cdots = 2^n
\end{equation}
states. Since no phonon appears more than once, the eigenvalues of 
the resulting $2^n \times 2^n$ matrix can be written as appropriate 
combinations of the eigenvalues of $n$ $2\times 2$ matrices. Even 
though the approach of CCFUS provides an elegant mathematical solution 
to the matrix diagonalization problem it ignores all the states where 
the same phonon appears more than once, e.g., the double-phonon states. 
\citeasnoun{nanda} and \citeasnoun{kruppa} pointed out that in some 
cases the coupling of a state like $(b_i^{\dagger})^2\ket{0}$ to the 
ground state can be stronger than the coupling of a state like 
$b_i^{\dagger} b_j^{\dagger} \ket{0}, i \ne j$. \citeasnoun{kruppa} 
considered quadrupole and octupole phonons, but included their 
double-phonon states as well diagonalizing the resulting $6\times 6$ 
matrix. \citeasnoun{nanda} excluded all multiple-phonon states, so 
for $n$ different types of phonons they numerically 
diagonalized an $(n+1) \times (n+1)$ matrix instead of analytically 
diagonalizing an $2^n \times 2^n$ matrix. The resulting simplified 
coupled channels code is named CCMOD \cite{nanda}. In CCMOD the weight 
factors are calculated at the position of the bare barrier, but 
the energy dependence of $R(E)$ in Eq. (\ref{6}) is taken into 
account using the prescription of \citeasnoun{neill}. Eigen-channel 
cross sections are again calculated using the Wong's formula. 

Many experimentalists use these simplified coupled channel codes. 
When using them it is important to remember the approximations 
discussed in the previous paragraphs. Some of these approximations 
(constant coupling, Wong's formula) lead to an overestimate of the 
cross section. Some of the ignored couplings (e.g., the double 
phonon states) may be very important for the dynamics of the 
analyzed system. We therefore recommend using these codes only for 
a qualitative understanding of the data and strongly encourage 
authors to use full coupled channel codes for any quantitative 
description. 

Finally one should point out that in the limit in which the 
intrinsic energies are small compared to 
the coupling interaction one can approximate  Eq. (\ref{mequation}) 
\begin{equation}
  \label{eq:zeroq}
  M_{nm} \sim F(r) \bra{n} G(\xi) \ket{m}.
\end{equation}
In this case it is not necessary to require the coupling to be 
constant. The transformation amplitude between the ground state and 
the eigen-channels labeled by $\xi$ is the ground state wave 
function yielding the transmission probability 
\begin{equation}
  \label{eq:zeroqt}
  T_{rm total} = \int d\xi |\psi(\xi)|^2 T [ E, V(r)+F(r)G(\xi)],
\end{equation}
where $T[ E, V(r)+F(r)G(\xi)]$ is the transmission probability for the 
potential $V(r)+F(r)G(\xi)$ calculated at energy $E$. 

\subsection{Path Integral Approach}
\label{IIIb}
 
An alternative formulation of the multidimensional quantum tunneling
is given by the path integral formalism \cite{ap}. 
For the Hamiltonian given in Eq. (\ref{29}) 
the propagator to go from an
initial state characterized by relative radial coordinate (the
magnitude of $\bf r$) $r_i$ and internal quantum numbers $n_i$ to a
final state characterized by the radial position $r_{f}$ and the
internal quantum numbers $n_f$ may be written as
\begin{equation}
K(r_f,n_f,T;r_i,n_i,0)=\int{\cal
D}\left[r(t)\right]e^{\frac{i}{\hbar}S(r,T)}W_{n_fn_i}(r(t),T),
\end{equation}
where $S(r,T)$ is the action for the translational motion and 
$W_{n_fn_i}$ is the propagator for the internal system 
along a given path of the translational motion:
\begin{equation}
W_{n_fn_i}(r,T)=\left\langle n_f\left| \hat U_{\rm int}
(r(t),T)\right|n_i\right\rangle.
\label{Weq}
\end{equation}
$\hat U_{\rm int}$ satisfies the differential equation
\begin{eqnarray}
i\hbar\frac{\partial\hat U_{\rm int}}{\partial t} &=&
\left[ H_0 + H_{\rm int} \right] \hat U_{\rm int}, \label{inteq}\\
\hat U_{\rm int}(t=0) &=& 1.
\end{eqnarray}
We want to consider the case where $r_i$ and $r_f$ are on opposite
sides of the barrier. In the limit when the initial and final states
are far away from the barrier, the transition amplitude is given by
the $S$-matrix element, which can be expressed in terms of the
propagator as \cite{ap}
\begin{eqnarray}
S_{n_f,n_i}(E) &=& -\frac{1}{i\hbar}\lim_{r_i\rightarrow\infty\atop
r_f\rightarrow-\infty}
\left(\frac{{\rm p}_i{\rm p}_f}{\mu^2}\right)^{\frac{1}{2}}{\rm exp}
\left[\frac{i}{\hbar}({\rm p}_fr_f-{\rm p}_ir_i)\right]\nonumber\\
&&
\int\limits_0^\infty dTe^{+iET/\hbar}K(r_f,n_f,T;r_i,n_i,0),
\label{smatrix}
\end{eqnarray}
where p$_i$ and p$_f$ are the classical momenta associated with $r_i$
and $r_f$. In heavy ion fusion we are interested in the transition
probability in which the internal system emerges in any final
state. For the $\ell$th partial wave, this is
\begin{equation}
T_\ell (E)=\sum_{n_f}|S_{n_f,n_i}(E)|^2,
\end{equation}
which becomes, upon substituting Eqs.\ (\ref{Weq}) and (\ref{smatrix}),
\begin{eqnarray}
T_\ell (E) &=& \lim_{r_i\rightarrow\infty\atop r_f\rightarrow-\infty}
\left(\frac{p_ip_f}{\mu^2}\right)\int\limits_0^\infty dT \exp
\left[{\frac{i}{\hbar}ET}\right]\int\limits_0^\infty\widetilde T \exp
\left[{-\frac{i}{\hbar} E\widetilde T}\right]\nonumber\\ &&\int{\cal
D}[r(t)]\int{\cal D}[\tilde r(\tilde t)] \exp \left[\frac{i}{\hbar}
(S(r,T)-S(\tilde r,\widetilde T))\right]\rho_M(\tilde r(\tilde
t),\widetilde T; r(t),T).
\label{influencefunc}
\end{eqnarray}
Here we have assumed that the energy dissipated to the internal system
is small compared to the total energy and taken p$_f$ outside the sum
over final states. We identified the two-time influence functional as
\begin{equation}
\rho_M(\tilde r(\tilde t),\widetilde T;r(t),T)
=\sum_{n_f}W^*_{n_f,n_i}(\tilde r(\tilde t);\widetilde T,0)
W_{n_f,n_i}(r(t);T,0).
\end{equation}

\noindent
Using the completeness of final states, we can simplify this expression 
to write
\begin{equation}
\rho_M(\tilde r(\tilde t),\widetilde T;r(t),T)
=\left\langle n_i\left| \hat U_{\rm int}^{\dagger}
({\tilde r}({\tilde t}),{\tilde T}) \hat U_{\rm int}
(r(t), T)\right|n_i\right\rangle.
\label{rhoeq}
\end{equation}
Eq. (\ref{rhoeq}) shows the utility of the influence functional method
when the internal system has symmetry properties. If the Hamiltonian
in Eq. (\ref{inteq}) has a dynamical or spectrum generating symmetry,
i.e., if it can be written in terms of the Casimir operators and
generators of a given Lie algebra, then the solution of
Eq. (\ref{inteq}) is an element of the corresponding Lie group
\cite{ap}.  Consequently the two time influence functional of
Eq. (\ref{rhoeq}) is simply a diagonal group matrix element for the
lowest-weight state and it can be evaluated using standard
group-theoretical methods. This is exactly the reason why the path
integral method is very convenient when the internal structure is
represented by an algebraic model such as the Interacting Boson Model.

Two-time influence functionals can be calculated exactly for only a
limited number of systems. One of these is a harmonic oscillator,
linearly coupled to the translational motion. In this case the
Hamiltonian is
\beq
H = -{\hbar^2\over {2\mu_0}}{\partial^2\over \partial
r^2}+V_0(r)+(a^{\dag }a+ {1\over 2})\hbar \omega+\alpha_0
f(r)(a+a^{\dag }),
\label{lincoup}
\eeq where $\mu_0 $ is the bare mass of the macroscopic motion and
$V_0(r)$ is the bare potential.  The $a^{\dag} $ ($a$), $\omega $, $m
$ and $f(r)$ are the creation (annihilation) operators of the
oscillator quanta, the frequency and the mass parameter of the
oscillator and the coupling form factor, respectively.  The quantity
$\alpha_0 = \left[ \hbar / 2 m \omega \right]^{1/2}$ represents the
amplitude of the zero-point motion of the harmonic oscillator.  The
two time influence functional $\rho_M$ reflects the effects of
coupling to the harmonic oscillator and is given by 
\beq \rho_M
\bigl({\widetilde r}({\tilde t}),{\widetilde T};r(t),T\bigr)=
\exp[-{i\over2}\omega(T-{\widetilde T})] \exp \bigl\lbrace -{\alpha_0^2
\over \hbar^2}(y_1+y_2+y_3)\bigr\rbrace
\label{ttime}
\eeq
with
\begin{eqnarray}
&y_1=&\int^T_0 dt \int^t_0ds ~f\bigl(r(t)
\bigr)f\bigl(r(s)\bigr)e^{-i\omega (t-s)}\nonumber\\
&y_2=&\int^{\widetilde T}_0 d{\tilde t} \int^{\tilde t}_0 d{\tilde s}
~f\bigl ({\widetilde r}({\tilde t})\bigr)f\bigl({\widetilde r}({\tilde
s})\bigr) e^{i\omega ({\tilde t}-{\tilde s})}\nonumber\\
&y_3=&-e^{i\omega ({\widetilde T}-T)}\int^T_0 dt ~f\bigl(r(t)\bigr)
e^{i\omega t}\int^{\widetilde T}_0 d{\tilde t} ~f\bigl({\widetilde r}
({\tilde t})\bigr)e^{-i\omega {\tilde t}}.
\label{3piece}
\end{eqnarray}
An exact calculation of the influence functional is possible in this
case because of the symmetry of the Hamiltonian under the
Heisenberg-Weyl algebra. 

\subsection{Adiabatic and Sudden Tunneling}

We can discuss the effects of couplings between nuclear structure and
translational motion in two limiting cases. The first case is the
sudden limit in which the energy levels of the internal system are
degenerate. The second case is the adiabatic limit in which the energy
of the first excited state of the system is very large so that the
internal system emerges in the ground state at the other side of the
barrier.
 
Several examples are worked out explicitly by \citeasnoun{ap}, to
which the reader is referred for further details. It can be shown that
in the sudden limit, the total transmission probability is given by an
integral of transmission probabilities for a fixed value of the
internal coordinate, with the weight of the integration given by the
distribution of the internal coordinate in its ground state.  For
example for the linearly-coupled harmonic oscillator as the excitation
energy gets smaller, $\omega \rightarrow 0$, the influence functional
of Eq. (\ref{ttime}) takes the form 
\beq \rho_M \bigl({\widetilde
r}({\tilde t}),{\widetilde T};r(t),T\bigr)=
\exp\left(-{\alpha^2_0\over 2\hbar^2}\bigl[\int^T_0dtf(r(t))-\int^
{\widetilde T}_0d{\tilde t}f({\widetilde r}({\tilde t}))\bigr]^2
\right).
\label{sudlin}
\eeq
Using the integral
\beq
\int^\infty_{-\infty}dxe^{-(ax^2+bx)}=\sqrt{{\pi\over a}}e^{b^2/4a}
\eeq
we can rewrite the influence functional as
\begin{eqnarray}
&\rho_M& \bigl({\widetilde r}({\tilde t}),{\widetilde T};r(t),T\bigr)
= \rho_{S} \bigl({\widetilde r}({\tilde t}),{\widetilde
T};r(t),T\bigr)\nonumber\\ &\equiv&{1\over \alpha_0\sqrt
{2\pi}}\int^\infty_{-\infty} d\alpha e^{-{1\over 2}({\alpha \over
\alpha_0})^2} \nonumber\\ &\times& \exp \left(-{i\alpha \over
\hbar}\bigl[\int^T_0dtf(r(t))-\int^ {\widetilde T}_0d{\tilde
t}f({\widetilde r}({\tilde t}))\bigr] \right)
\end{eqnarray}
where the lower suffix $s$ stands for sudden tunneling.  Inserting
this expression into Eq. (\ref{influencefunc}) one obtains
\begin{equation}
T (E) = {\frac{1}{\sqrt {2 \pi}}} \int_{- \infty}^{+\infty} dx e^{-x^2
/ 2} T_0 (E, V_0(r) + x \alpha_0 f(r)),
\label{zero}
\end{equation}
where $T_0 (E, V_0(r) + x \alpha f(r))$ is the probability of
tunneling through the one-dimensional barrier $V_0(r) + x \alpha f(r)$
at energy $E$. This expression is known as the zero point motion
formula and was first derived by Esbensen \cite{zpm1}. This result can
also be derived either using the coupled-channels formalism (cf. 
Section \ref{IIIanew}) or using Green's functions \cite{yoram}.  
Similarly, if
the translational motion couples to a very slow rotational motion
through a coupling Hamiltonian given by 
\beq 
H_{int}=\sqrt{{5\over 4\pi}} \beta P_2(cos\theta) f(r)
\label{bill}
\eeq 
then the net tunneling probability is obtained by first
calculating the tunneling probability for a fixed orientation of the
principal axis of the deformed nucleus, and then taking average over
all orientations 
\beq 
T(E) = \int^1_0 d\cos\theta T_0 \left( E,V_0(r)+ 
\sqrt{{5\over 4\pi}}\beta 
P_2(cos\theta) f(r) \right).  
\label{deform}
\eeq 
This formula was first derived by \citeasnoun{chase} in the study of
scattering of rotational nuclei in the sudden approximation.
Systematics of the fusion cross sections of $^{16}$O with a series of
Sm isotopes, ranging from vibrational to rotational, was first given
by \citeasnoun{stok2} using Eq. (\ref{deform}). 

In actual calculations, both vibrational and rotational excitations
are truncated at finite excited states. In these cases, the Hermite
and the Gauss integrals in Eqs. (\ref{zero}) and (\ref{deform}) are
replaced by the Hermite and the Gauss quadratures, respectively
\cite{geo}. Zero point motion formulae then have a simple geometric
interpretation: In this approximation fusion of a deformed nucleus
with a finite number (N) of levels can be described by sampling N
orientations with their respective weights: 
\beq
\sigma (E) = \sum_{i=1}^N \omega_i \sigma (E, V_0(R) + \lambda_i f(r)),
\eeq
where $\sum_{i=1}^N \omega_i =1$. 
For example, in a two-level system, the orientations
$\theta_1=70.12^o$ and $\theta_2=30.55^o$ contribute with weight
factors $\omega_1=0.652$ and $\omega_1=0.348$, respectively, as
illustrated in Figure \ref{Fig4a}.

When both target and projectile are heavy and deformed one needs a
model describing macroscopic potential energy surfaces for arbitrarily
oriented, deformed heavy ions. Such a model describing completely
general configurations of two separated nuclei is given by
\citeasnoun{moller}. 

In the adiabatic limit, i.e., that of slow tunneling, one can
introduce \cite{ap} an $\Omega / \omega$ expansion, $\omega$ and
$\Omega$ being the frequencies of the internal motion and the
tunneling barrier, respectively. In this limit the effects of the
coupling to internal degrees of freedom can be represented in terms of
energy-independent potential renormalization \cite{ap,muller} :
\begin{equation}
V_0(r) \rightarrow V_{ad} = V_0(r) - \alpha_0^2 f^2(r) / \hbar \omega,
\label{potren}
\end{equation}
and a mass renormalization \cite{massrenorm} \beq 
\mu_0 \rightarrow
\mu_{\rm ad} = \mu_0 + 2{\alpha_0^2\over \hbar \omega} {1\over
\omega^2}\left({df\over dr}\right)^2.
\label{massren}
\eeq 
This limit is very closely related to the situation investigated by
\citeasnoun{leg} in their study of multidimensional quantum
tunneling. In the Caldeira-Leggett formalism the term which
renormalizes the potential in Eq. (\ref{potren}) is added to the
Lagrangian as a counter-term. Note that the correction term to 
the potential is the well-known polarization potential and 
the correction term to the mass
is the cranking mass. Hence appropriate generalizations of 
Eqs. (\ref{potren}) and (\ref{massren}) hold  
not only for a linearly-coupled oscillator, but for any system
\cite{massrenorm}.
 Effects of the polarization potential on the 
subbarrier fusion cross section were elucidated by 
\citeasnoun{tanimura}. 

Typically adiabatic barrier alone overpredicts the transmission 
probability. We can demonstrate this for the constant coupling case 
($\mu_{\rm ad} = \mu_0$) using the exact coupled-channels result 
of Eq. (\ref{simcc})
\begin{equation}
\label{simccprime}
  T_{\rm cc}(E) = \sum_{n=0}^{\infty} \frac{1}{n!} 
\left( \frac{\alpha_0 f}{\hbar \omega} \right)^{2n}  
\exp(-\alpha_0^2f^2/(\hbar\omega)^2) T \left( E-n\hbar \omega + 
\frac{\alpha_0^2 f^2}{\hbar\omega} \right). 
\end{equation}
Inserting the inequality
\begin{equation}
  \label{eq:adiien}
  T \left( E-n\hbar \omega + 
\frac{\alpha_0^2 f^2}{\hbar\omega} \right) \le T \left( E + 
\frac{\alpha_0^2 f^2}{\hbar\omega} \right). 
\end{equation}
into Eq. (\ref{simccprime}) we get
\begin{equation}
  \label{eq:adicc}
  T_{\rm cc}(E) \le T_{\rm adiabatic}(E).
\end{equation}

We should emphasize that in the adiabatic limit both the potential 
and the mass renormalization should be considered together. Many 
authors refer to using only the adiabatic potential as the 
adiabatic limit. While for constant coupling it is true that 
$\mu_{\rm ad} 
= \mu_0$, in most cases of interest to nuclear fusion the 
coupling form factors rapidly change near the barrier and the 
difference between the adiabatic mass and the bare mass can get 
very large. In these cases, using the adiabatic correction to the 
mass in addition to the adiabatic potential, 
even though the former is in the next order in $1/\omega$ as compared 
to the latter, can significantly reduce the transmission probability 
to values below the exact coupled-channels result \cite{massrenorm}. 

Adiabatic and sudden approximations are very useful for obtaining 
analytical results which provide a conceptual framework for 
understanding the fusion process. For deformed nuclei, where 
the excitation energies are very low, sudden approximation provides 
a reasonably good description of the data. In case of 
rotation-vibration coupling, sudden approximation can be utilized 
to reduce the size of the channel coupling. 

If the excitation energies of the internal system are large the 
sudden approximation tends to overestimate the tunneling 
probability at energies well below the barrier. 
Indeed \citeasnoun{finiteex} showed that at below 
the barrier energies sudden approximation provides 
an upper limit to the tunneling probability. 
Using the path integral approach 
\citeasnoun{taki95b} showed that, for a nearly degenerate system, 
the finite excitation energy leads to a multiplicative dissipation 
factor which reduces the barrier penetrability estimated in the 
sudden limit. This latter result also provides a good example of 
the utility of the path integral method. As we mentioned earlier 
for Hamiltonians written 
in terms of the generators of a Lie algebra path integral approach 
is especially convenient as integrals over paths become integrals 
over the group manifold and thus are amenable to the standard 
group theoretical techniques. An example for this is the treatment 
of the nuclear structure effects in subbarrier fusion with the 
interacting boson model discussed in Section \ref{IV}. Finally 
coupled channel techniques are not practical as the number of 
channels gets very large, e.g., for tunneling problems at finite 
temperature. On the other hand for finite temperature problems 
path integral techniques can easily be applied (see e.g., 
\citeasnoun{abe-kun} where particle decay from a hot compound 
nucleus is investigated and the thermal fluctuation of the nuclear 
surface is shown to amplify the dynamical effect). 

\subsection{Intermediate Cases - Dynamical Norm Method}
\label{IIID}

In the intermediate cases between the sudden and the adiabatic
tunneling, the effects of the environment is not straightforward to
illustrate in simple physical terms. The dynamical norm method
\cite{brr,dynamo} is a technique introduced to give an intuitive
understanding of the effects of the environment in the intermediate
cases.  Defining the adiabatic basis by
\beq \bigl[
H_0(\xi)+H_{int}(\xi,r(t))\bigr]\phi_n(r(t),\xi)  \equiv
h(r(t),\xi)\phi_n(r(t),\xi) =\epsilon_n(r(t))\phi_n(r(t),\xi) .  
\eeq
In the same spirit as the WKB approximation for a potential model, for
the total wave function one can take the ansatz 
\beq
\Psi(r,\xi)=\Phi(r,\xi)\left({{dW}\over{dr}}\right)^{-1/2}e^{i\epsilon
W(r)/\hbar}.  \eeq 
Here the parameter $\epsilon^2 $ is $1$ and $-1$ in classically
allowed and in classically forbidden regions, respectively and the
action obeys the Hamilton-Jacobi equation 
\beq
{{\epsilon^2}\over{2\mu_0}}\left({{dW}\over{dr}}\right)^2+V_{ad}(r)=E
\eeq 
where $V_{ad}$ is the adiabatic potential defined by 
\beq 
V_{ad}(r)=V_0(R)+\epsilon_0(r).  \eeq 
One can then derive an approximate expression where the net tunneling
probability is given by the product of the tunneling probability
through the adiabatic potential barrier and a multiplicative factor
which represents the non-adiabatic effect of the intrinsic degrees of
freedom :
\begin{eqnarray}
\label{yenidyn}
T&=&T_0(E,V_{ad})\cdot {\cal N}(\tau_b)\nonumber\\
{\cal N}(\tau)&=&\int
d\xi \vert\Phi\bigl(r(\tau),\xi\bigr)\vert^2, 
\end{eqnarray}
where $\tau_b $ is the time when the tunneling process is completed.
Note that in this approximation the transmission probability in 
Eq. (\ref{yenidyn}) is calculated with the bare mass, not the 
renormalized mass of Eq. (\ref{massren}). 
One can easily show that ${\cal N} \leq 1$. This means that the
transmission probability calculated with the adiabatic potential 
is always greater than the actual transmission probability. The
deviation of ${\cal N}$ from 1 gives the measure of non-adiabaticity
of the tunneling process.

One way to determine the dynamical norm factor, $\cal N$, is to first
expand $\Phi (r,\xi)$ in the basis of the adiabatic states
$\phi_n(r(\tau), \xi)$
\beq
\Phi(r,\xi)=\sum_na_n(\tau)\cdot \phi_n(r(\tau),\xi)
\eeq
and then determine the expansion coefficients $a_n$ by solving the
coupled-channels equations
\begin{eqnarray}
{\dot a}_n(\tau)-{\dot r}(\tau)\sum_{m\not= n}a_m(\tau)
{1\over \epsilon_n(r(\tau))-\epsilon_m(r(\tau))}
\nonumber \\
<\phi_n\vert{\partial \tilde h\over \partial r}\vert \phi_m>
=-{1\over \hbar}{\tilde \epsilon_n(r(\tau))}a_n(\tau)
\label{haginonorm}
\end{eqnarray}
with ${\tilde \epsilon}_n=\epsilon_n(r(\tau))-\epsilon_0(r(\tau))$.
Note that Eq. (\ref{haginonorm}) is correct to describe a classically
forbidden region. One needs to modify it with $\epsilon $ in order to
describe a classically allowed region.  As an example, one can
consider the linearly-coupled oscillator with the coupling form factor
given by $f(r)=cr$. Assuming that the tunneling path is given by
$t(\tau)=R_0\sin(\Omega\tau)$, $R_0$ being the length of the tunneling
region, we obtain 
\beq 
{\cal N}(\tau_b) \sim \exp \left\{-{\pi\over
2} \left( {c\alpha_0R_0\over \hbar\omega}\right)^2 
{\Omega\over \omega}\right\}
\label{dynamon}
\eeq
Eq. (\ref{dynamon}) shows that the adiabaticity of the tunneling
process is governed not only by a parameter $\Omega / \omega$, but
also on the coupling strength and on the length of the tunneling
region. \citeasnoun{dynamo} applied the dynamical norm method to 
spontaneous fission of $^{234}$U.

\section{Description of Nuclear Structure Effects by the Interacting
Boson Model}
\label{IV}

An algebraic nuclear structure model significantly simplifies
evaluation of the path integral.  The Interacting Boson Model (IBM) of
Arima and Iachello \cite{ibm} is one such model which has been
successfully employed to describe the properties of low-lying
collective states in medium-heavy nuclei. In this section, attempts to
use the IBM in describing nuclear structure effects on fusion are
reviewed.  The path integral formulation of this problem, as sketched
in the next section, requires analytic solutions for the nuclear wave
functions. In the first attempt to use the IBM to describe nuclear
structure effects in subbarrier fusion, the SU(3) limit of the IBM was
employed \cite{ibm1}. However, the SU(3) limit corresponds to a rigid
nucleus with a particular quadrupole deformation and no hexadecapole
deformation, a situation which is not realized in most deformed
nuclei. Thus analytic solutions away from the limiting symmetries of
the IBM are needed for realistic calculations of subbarrier fusion
cross sections.

In a parallel development, a 1/N expansion was investigated
\cite{serdar1,serdar3,serdar4} for the IBM which provided 
analytic solutions for a
general Hamiltonian with arbitrary kinds of bosons. This technique
proved useful in a variety of nuclear structure problems where direct
numerical calculations are prohibitively difficult. Later it was
applied to medium energy proton scattering from collective nuclei
\cite{serdar2} in the Glauber approximation, generalizing the earlier
work done using the SU(3) limit \cite{joe}. Using the 1/N expansion
technique in the path integral formulation of the fusion problem
\cite{ibm2} makes it possible to go away from the three symmetry
limits of the IBM. In particular arbitrary quadrupole and hexadecapole
couplings can be introduced.

\subsection{Linear Coupling}

We take $H_{\rm int}$ in Eq. (\ref{29}) 
to be of the form of the most general one-body
transition operator for the IBM,
\begin{equation}
H_{\rm int}= \sum_{kj\ell}\alpha_{kj\ell}(r)\left[b^\dagger_j\tilde
b_\ell\right]^{(k)} \cdot Y^{(k)}(\hat{\bf r}),
\label{linearIBM}
\end{equation}
where the boson operators are denoted by $b_\ell$ and $b^\dagger_j$.
The $k$ sum runs over $k=2,4,\ldots2\ell_{\rm max}$. Odd values of $k$
are excluded as a consequence of the reflection symmetry of the
nuclear shape, and the $k=0$ term is already included in the bare
potential $V_0(r)$. The form factors $\alpha_{kj\ell}(r)$ represent
the spatial dependence of the coupling between the intrinsic and
translational motions. The interaction term given in
Eq. (\ref{linearIBM}) is an element of the $SU(6)$ algebra for the
original form of the Interaction Boson Model with s and d bosons and
is an element of the $SU(15)$ algebra when g bosons are included as
well \cite{ibm}.

To simplify the calculation of the influence functional, we use the
no-Coriolis approximation. We first perform a rotation at each instant
to a frame in which the $z$-axis points along the direction of
relative motion. Neglecting the resulting centrifugal and Coriolis
terms in this rotating frame is equivalent to ignoring the angular
dependence of the original Hamiltonian.  In this approximation, the
coupling form factors become independent of $\ell$ and only $m=0$
magnetic sub-states of the target are excited
\cite{nob1,nob2,nob3}. For heavy systems the neglected centrifugal and
Coriolis forces are small. We take the scattering to be in the $x$-$y$
plane. Then making a rotation through the Euler angles $\hat{\bf
b}=(\phi,\pi/2,0)$, we can write the full Hamiltonian as the rotation
of a simpler Hamiltonian depending only on the magnitude of $\bf r$
\begin{equation}
H=R(\hat{\bf b})H^{(0)}(r)R^\dagger(\hat{\bf b}).\label{ham}
\end{equation}
Since in Eq. (\ref{29}) $H_0$ and $H_t=H_k+V_0(r)$ are rotationally
invariant, $H_{\rm int}$ is the only term whose form is affected by
the transformation. Hence we introduce the rotated interaction
Hamiltonian $H_{\rm int}^{(0)}(r)$, given by
\begin{eqnarray}
H_{\rm int} &=& R(\hat{\bf b})H_{\rm int}^{(0)}(r)R^\dagger(\hat{\bf
b}),\\ H_{\rm int}^{(0)}(r) &=& \sum_{j\ell m}\phi_{j\ell
m}(r)b^\dagger_{jm}b_{\ell m},\\ \phi_{j\ell m}(r) &=&
(-)^m\sum_k\sqrt{\frac{2k+1}{4\pi}}\langle jm\ell
-m|k0\rangle\alpha_{kj\ell}(r).
\end{eqnarray}
If we assume now that the form factors $\alpha_{kj\ell}(r)$ are all
proportional to the same function of $r$ then the Hamiltonian $H_{\rm
int}^{(0)}$ commutes with itself at different times and hence we can
write the two-time influence functional as
\begin{equation}
\rho_M=\left\langle n_i\left|\exp
\left(\frac{i}{\hbar}\int_0^{\widetilde T} dtH_{\rm int}^{(0)}(\tilde
r(t))\right) \exp\left(-\frac{i}{\hbar}\int_0^T dtH_{\rm
int}^{(0)}(r(t))\right)\right|n_i\right\rangle
\end{equation}
in the degenerate spectrum limit. 

Since the exponents of the two operators in the influence functional
commute, $\rho_M$ becomes the matrix element of an SU($6$)
transformation between SU($6$) basis states, in other words it is a
representation matrix element for this group and can easily be
calculated using standard techniques. The two-time influence
functional for the sd-version of the IBM was calculated by
\citeasnoun{ibm2} and, for the particular case of SU(3) limit, by
\citeasnoun{ibm1}.

\subsection{Higher-Order Couplings}
\label{IVb}

Up to this point, we have utilized only a first-order coupling between
nuclear states and translational motion. Alternatively, one can
include the effects of coupling to all orders. This can be achieved by
exploiting the symmetry properties of the resolvent operator directly
without utilizing its path integral representation. Such a Green's
function approach has also been used to study quantum tunneling in a
heat bath \cite{yoram}.

To include the effects of couplings to all orders, the interaction
Hamiltonian in Eq. (\ref{29}) is written as
\begin{equation}
H_{int}({\bf r}, {\xi}) + V_0(r) = V_{\rm Coul}({\bf r}, {\xi}) +
V_{\rm nuc}({\bf r}, {\xi}),
\end{equation}
where the Coulomb part is 
\begin{eqnarray}
V_{\rm Coul}({\bf r}, {\xi})&=&\frac{Z_1Z_2e^2}{r}(1 +
\frac{3}{5}\frac{R_1^2}{r^2}\hat O)~~ (r>R_1), \nonumber\\
                            &=&\frac{Z_1Z_2e^2}{r}(1 + 
\frac{3}{5} \frac{r^2}{R_1^2}\hat O)~~  (r<R_1).
\label{coulombi}
\end{eqnarray}
The nuclear part is taken to have Woods-Saxon form,
\begin{equation}
V_{\rm nuc}({\bf r}, {\xi}) = -V_0\left(1 + \exp \left({r -R_0 - 
R_1{\hat O}({\hat{\bf r}, {\xi}})\over a}\right)\right)^{-1}. 
\label{nucleari}
\end{equation}
In Eqs. (\ref{coulombi}) and (\ref{nucleari}), $R_0$ is the sum of the
target and projectile radii and $R_1$ is the mean radius of the
deformed target. $\hat O$ is a general coupling operator between the
internal coordinates and the relative motion
\begin{equation}
\hat O = \sum_ {k} v_k T^{(k)}({\xi}) \cdot Y^{(k)}(\hat{\bf r}) .
\end{equation}
The coefficients $v_k$ represent the strengths of the various
multipole transitions in the target nucleus.  In the standard IBM with
s and d bosons, the only possible transition operators have $k =
0,2,4,\dots$ (odd values being excluded as a consequence of the
reflection symmetry of the nuclear shape). The monopole contribution
is already included in the Woods-Saxon parameterization and so is not
needed. The quadrupole and hexadecapole operators are given by
\begin{eqnarray}
T^{(2)}&=&[s^{\dagger}{\tilde d}+d^{\dagger}s]^{(2)} + \chi
[d^{\dagger}{\tilde d}]^{(2)}, \label{quadeq}\\
T^{(4)}&=&[d^{\dagger}{\tilde d}]^{(4)}. 
\end{eqnarray}
We adopt the ``consistent-Q'' formalism of Casten and Warner
\cite{warnercasten}, in which $\chi$ in Eq. (\ref{quadeq}) is taken to
be the same as in $H_{IBM}$ (fitted to reproduce the energy level
scheme and the electromagnetic transition rates of the target nucleus)
and is thus not a free parameter.

In the previous section, we used the usual approximation in which the
nuclear potential of Eq. (\ref{nucleari}) is expanded in powers of the
coupling, keeping only the linear term (cf. Eq. (\ref{linearIBM})). In
order to calculate the fusion cross section to all orders we consider
the resolvent operator for the system
\begin{equation}
G^{+}(E) = \frac{1}{E^{+}-H_t-H_{IBM}({\xi})-H_{\rm int}(r,
\hat O)}.
\label{denklem2}
\end{equation}
The basic idea is to identify the unitary transformation which
diagonalizes the operator $\hat O$
\begin{equation}
\hat O_d = {\cal U} \hat O {\cal U}^{\dagger}
\end{equation}
in order to calculate its eigenvalues and eigenfunctions 
\begin{equation}
\hat O_d | n \rangle = \zeta_n | n \rangle .
\end{equation}
Assuming the completeness of these eigenfunctions 
\begin{equation}
\sum_n | n \rangle \langle n| =1 
\end{equation}
one can write the matrix element of the resolvent as 
\begin{eqnarray}
&\langle& \xi_f , r_f \> | \> G^{+}(E) \> | \> \xi_i , r_i \> \rangle
\nonumber \\ & = & \langle \xi_f , r_f | {\cal U}^{\dagger} {\cal U}
\left[ E^{+}-H_t-H_{\rm int}(r, \hat O) \right]^{-1} {\cal
U}^{\dagger} {\cal U} | \xi_i , r_i \rangle \nonumber\\ & = & \langle
\xi_f , r_f | {\cal U}^{\dagger} \left[ E^{+}-H_t-H_{\rm int}(r, \hat
O_d) \right]^{-1} \sum_n | n \rangle \langle n| {\cal U} | \xi_i , r_i
\rangle \nonumber\\ & = & \sum_n \langle \xi_f | {\cal U}^{\dagger} |
n \rangle \langle n| {\cal U} | \xi_i \rangle \langle r_f | G^{+}_n |
r_i \rangle
\label{resolvent}
\end{eqnarray}
where 
\begin{equation}
G^{+}_n(E) = \frac{1}{E^{+}-H_t-H_{\rm int}(r, \zeta_n)}.
\label{denklem}
\end{equation}
To derive this result we ignore the excitation energies in the target
nucleus. This corresponds to setting the term $H_{IBM}$ to be zero in
Eqs. (\ref{denklem2}) and (\ref{denklem}).  In this case, the
$G^{+}_n(E)$ given in Eq. (\ref{resolvent}) is the resolvent operator
for one-dimensional motion in the potential $H_{\rm int}(r, \zeta_n)$,
the fusion cross section of which can easily be calculated within the
standard WKB approximation. The total cross section can be calculated
by multiplying these eigen-channel cross sections by the weight factors
indicated in Eq. (\ref{resolvent}). The calculation of the matrix
element $\langle n| {\cal U} | \xi_i \rangle$ within the IBM is
straightforward \cite{ibm3,ibm4,ibm5}. It is also possible to
generalize the previous formalism to include arbitrary kinds of bosons
in the target nucleus and investigate whether g bosons have any
discernible effects on subbarrier fusion reactions.  One finds that
\cite{ibm5} except for slight differences in the barrier distributions
(which can be made even smaller by fine tuning the coupling
strengths), there are no visible differences between the {\it sd} and
{\it sdg} model predictions. The similarity of the results implies
that subbarrier fusion probes the overall coupling strength in nuclei,
but otherwise is not sensitive to the details of the nuclear
wave functions. In this sense subbarrier fusion, which is an 
inclusive process, is in the same
category as other static quantities (energy levels, electromagnetic
transition rates), and does not seem to constitute a dynamic probe of
nuclei, in contrast to exclusive processes such as proton scattering.

Using this formalism, a systematic study of subbarrier fusion of
$^{16}$O with rare earth nuclei became possible. Fusion cross sections
for the reactions $^{16}$O + $^{144,148,154}$Sm and $^{16}$O +
$^{186}$W were measured by the Australian National University group
\cite{wei,lei93,lemmon,morton,lei95}. The angular momentum 
distributions for
$^{16}$O + $^{154}$Sm was measured by \citeasnoun{sm154}, and for
$^{16}$O + $^{152}$Sm by \citeasnoun{wuosma}. Those for $^{16}$O +
$^{144,148}$Sm were deduced by \citeasnoun{baba1}. \citeasnoun{ibm4}
fit the existing data on vibrational and rotational nuclei with a
consistent set of parameters which are then used to predict the
cross-section and $<\ell>$-distributions.  Figure \ref{Fig5} compares
those data with the cross section calculations of \citeasnoun{ibm4}
and the angular momentum distribution calculations of
\citeasnoun{ibm6}.

Finally one should mention that the effects of anharmonicities both in
nuclear spectra \cite{zamfir} and in the vibrational coupling in
subbarrier fusion \cite{hindenew,mord} recently attracted some
attention. \citeasnoun{arima} pointed out that the $U(5)$ symmetry
limit of the IBM should exhibit in the spectra anharmonicities similar
to the geometric anharmonic vibrator model of \citeasnoun{piza}. This
assertion was later explicitly confirmed \cite{ani}. This feature of
the IBM makes it possible to discuss the effects of anharmonicities in
the vibrational coupling in subbarrier fusion using the $U(5)$ limit
\cite{u5}.

\section{Comparison of Current Theory with Data}
\label{V}

\subsection{Status of Coupled-Channels Calculations}

\citeasnoun{diG91} presented a systematic analysis of fusion cross
sections and average angular momenta for fourteen different systems
using a barrier penetration model that includes coupling to inelastic
channels. They concluded that model predictions explain data well for
light and asymmetric systems whereas large discrepancies exist for
large symmetric systems. Indeed for light and asymmetric systems the
basic premise of the coupled-channels calculations is justified: For
these systems the repulsive Coulomb potential is relatively weak and
the tail of the attractive nuclear potential has sufficient strength
to ``turn it around'' to form the potential barrier. The barrier is
thus formed at a rather large nuclear separation, long before two
nuclear surfaces start touching. Consequently, as the system
penetrates the barrier individual nuclei preserve their character and
one can talk about coupling of the states in the target nuclei to the
quantum tunneling process. The inversion procedure of \citeasnoun{bkn}
demonstrated that even when there are isotopic differences in the
fusion cross section \cite{wu}, one can still describe the quantum
tunneling with a one-dimensional {\em effective} potential for very
light systems. On the other hand, for heavier and more symmetric
systems other effects not explicitly included in the coupled-channels
calculations, such as neck formation may play an important role.

Many experimental groups use simplified coupled-channels codes such as
CCFUS \cite{dasso2}, CCDEF \cite{CCDEF}, CCMOD \cite{nanda}, or the
IBM-based models \cite{ibm3,ibm4,ibm5}. As we mentioned earlier it is
worthwhile to keep in mind that although these codes are quite
adequate for qualitative comparisons, they may be making a number of
assumptions, such as ignoring the radial dependence of the coupling
form factor, excitation energies and/or higher-order couplings. Before
one makes a quantitative statement, it may be sagacious to check what
the approximations are and if the observed discrepancies with the data
is a result of these simplifications. In Section \ref{IVb}, we
explicitly demonstrated the effect of the higher order couplings.

The crucial ingredient of the coupled-channels calculations is to
identify relevant degrees of freedom and to model the appropriate
Hamiltonian. Even the choice of the optical potential should be
scrutinized. For example, a recent survey \cite{brandan} of the
knowledge of the optical potential between even much simpler systems,
such as two light ions, indicate that many anomalies needed to be
resolved before a good theoretical understanding of the elastic
scattering data can be achieved. For heavier systems, from fits to
elastic scattering data at energies near the barrier, the optical
potential was shown to have a strong energy dependence, known as
``threshold anomaly'' \cite{satch}. \citeasnoun{optic} pointed out
that the dispersion relation between the real and imaginary parts of
the optical potential should be used in regions where the absorption
varies rapidly with energy such as near and below the barrier. The
source of the energy dependence can be either channel coupling or the
non-locality of the exchange contribution
\cite{marco1}. \citeasnoun{marco2} compared non-local effects and
coupled channels calculations in simple models of nuclear fusion.

The very first coupled-channels calculations for heavy-ion fusion
assumed a linear coupling to quadrupole or octupole surface vibrations
and quadrupole deformations. As more precise data became available,
the significance of the hexadecapole deformations \cite{rbrown},
neutron transfer \cite{broglia}, coupling of multi-phonon states
\cite{nob1,kruppa}, and higher-order couplings \cite{ibm3} emerged.
In the rest of this section we discuss representative data
illustrating these effects.

\subsection{Nucleon Transfer}
\label{nuctrans}

Another interesting question is the effect of the nucleon transfer on
subbarrier fusion (cf. Section \ref{IIIanew}).  
In particular the role of transfer channels with
positive Q-values has been emphasized
\cite{broglia,broglia1,esbensen1}. Algebraic models, such as that
described in Section \ref{IV}, at present do not include the effects
of nucleon transfer. It is now experimentally possible to observe up
to six-nucleon transfer at subbarrier energies \cite{jiang}. Hence in
the near future systematic studies including nucleon transfer
reactions, fusion, and elastic scattering may be possible.

The effect of nucleon transfer on fusion can be illustrated for
example by considering fusion reactions between different Ni
isotopes. Indeed these were the pioneering experiments of
\citeasnoun{beck2} where the enhancement of subbarrier fusion cross
sections was first observed. These cross sections were later measured
by \citeasnoun{schicker} and more recently by \citeasnoun{ack96}. The
cross sections measured by \citeasnoun{ack96} for the
$^{58}$Ni$+^{64}$Ni and $^{64}$Ni$+^{64}$Ni systems are displayed in
Figure \ref{Fig6b}, where the energies are normalized to the height of
the s-wave potential barrier. One sees a discernible enhancement for
the $^{58}$Ni$+^{64}$Ni system over the $^{64}$Ni$+^{64}$Ni system. In
the symmetric system $^{64}$Ni$+^{64}$Ni there are no transfer
channels with positive Q-values, and only those channels describing
inelastic excitations need to be included. On the other hand, in the
$^{58}$Ni$+^{64}$Ni system there is an additional coupling of the
transfer channel $^{64}$Ni($^{58}$Ni,$^{60}$Ni)$^{62}$Ni with a
Q-value of $Q=+3.9$ MeV. These additional channels increase the
$^{58}$Ni$+^{64}$Ni cross section (cf. the discussion in Section 
\ref{IIIanew}). 

Signatures of positive Q-value transfer reactions can also be
identified in fusion barrier distributions. By comparing barrier
distributions for $^{16}$O$+^{144}$Sm and$^{17}$O$+^{144}$Sm reactions
\citeasnoun{morton} showed that the effect of the neutron-stripping
channel in the second reaction is evident in the barrier distribution.

In fusion reactions of identical nuclei there are a number of
interesting effects magnified by the existence of only even partial
waves. For example, the fusion cross sections have an oscillatory
structure as a function of energy \cite{poffe}. Furthermore, elastic
transfer plays an important role in such collisions
\cite{oer}. \citeasnoun{chris} showed that even in cases where no
oscillatory structure is visible in cross section, there still is a
signature of the elastic transfer in the barrier distributions.

\subsection{Probing Asymmetry Effects}

One way to probe the effects of asymmetry of the system other than
nucleon transfer is to measure fusion cross sections and average
angular momenta for different systems leading to the same compound
nucleus. Such a measurement was recently performed \cite{ack96} for
the systems $^{28}$Si$+^{100}$Mo and $^{64}$Ni$+^{64}$Ni leading to
the compound nucleus $^{128}$Ba. These measurements complement a
previous measurement for the system $^{16}$O$+^{112}$Cd
\cite{ack94}. They find that both fusion cross sections as well as
average angular momenta can be explained by coupled channels
calculations. For the $^{28}$Si$+^{100}$Mo system including lowest 
$2^+$ and $3^{-}$ states of both the target and the projectile in 
the coupled-channels calculation improves the agreement between 
theory and data, as compared to the  no-coupling limit, but is not 
sufficient to reproduce the data. One needs to include an additional 
channel with the Q-value of the two-neutron pick-up reaction to 
bring the data and theory into agreement. The data for the 
symmetric system $^{64}$Ni$+^{64}$Ni, where no transfer channels 
with positive Q-values are present, are already well reproduced only 
with coupling to the inelastic channels. Hence the data of 
\citeasnoun{ack96} also provide evidence for the influence of the
two-nucleon transfer channels with positive Q-values on the fusion
probabilities. 

Studies of transfer channel coupling and entrance channel effects for
the near and subbarrier fusion was also given by \citeasnoun{prasa}
for the systems $^{46}$Ti $+^{64}$Ni, $^{50}$Ti $+^{60}$Ni, $^{19}$F
$+^{93}$Nb, and by \citeasnoun{charlop} for the systems
$^{28}$Si$+^{142}$Ce, $^{32}$S$+^{138}$Ba, and
$^{48}$Ti$+^{122}$Sn. These authors also report no significant
entrance channel effects except that the positive Q value for
two-neutron pickup shows up as an additional enhancement in the
$^{46}$Ti $+^{64}$Ni system.

\subsection{Signatures of Nuclear Vibrations}

One relatively unexplored aspect of subbarrier fusion is searching for
signatures of nuclear vibrations. \citeasnoun{lei95} confirmed the
effects of vibrational coupling in the $^{16}$O+$^{144}$Sm system.  To
search for signatures of nuclear vibrations \citeasnoun{stef4}
measured the fusion cross section for $^{32,36}$S$+^{110}$Pd
systems. $^{110}$Pd is a vibrational nucleus whose
two-quadrupole-phonon triplet is well known. Simplified
coupled-channels calculations for these systems are performed by
\citeasnoun{stef4} and by \citeasnoun{deweerd}. \citeasnoun{stef4},
using the method of \citeasnoun{kruppa}, assumed a constant coupling
explicitly including the finite Q value of the coupled channels.  In
these experiments it was also observed that the cross section for the
$^{32}$S$+^{110}$Pd system is greatly enhanced because of the
two-neutron transfer channel.

Since $^{110}$Pd lies between $U(5)$ and $SO(6)$ symmetry limits of
the IBM it cannot be described analytically. \citeasnoun{deweerd}
first numerically calculated quadrupole matrix elements between
different states using the PHINT code \cite{phint} and then
numerically obtained eigenvalues and the associated weights
(cf. Eq. (\ref{resolvent})). His result for the barrier distribution
of the $^{36}$S$+^{110}$Pd system is displayed in Figure \ref{Fig6c}
along with the data. While one-phonon space clearly fails in
describing the barrier distribution, the agreement with data
successively improves as one includes more phonons in the
calculation. This drastic change in the barrier distribution for
different numbers of phonons was also noted by \citeasnoun{stef4} for
this system and by \citeasnoun{stef3} for the $^{58}$Ni$+^{60}$Ni
system.

\subsection{Effect of Non-linear couplings}

An important component in the theoretical description of the
subbarrier fusion data is the effect of non-linear couplings. These
effects were discussed by \citeasnoun{ibm3} using the IBM in the limit
of zero excitation energy (cf. Section \ref{IVb}). For nuclear surface
vibrations the excitation energies cannot be neglected in most cases,
and one has to solve full coupled-channels equations. These
calculations can be very involved, and consequently they were carried
out by very few groups.

\citeasnoun{esbensen} expanded the coupling potential
(cf. Eq. (\ref{nucleari})) up to second order with respect to the
deformation parameter obtaining a good agreement between their
calculations and data for fusion cross sections between different
nickel isotopes. The quadratic coupling approximation was shown to
describe well fusion cross sections and angular momentum distributions
for the $^{58,64}$Ni$+^{92,100}$Mo reactions
\cite{rehm}. Coupled-channels calculations including coupling to all
orders and the finite excitation energy of nuclear surface vibrations
were performed for the $^{58}$Ni$+^{60}$Ni reaction \cite{stef3}.

In Figure \ref{Fig7} the calculation of \citeasnoun{hag97} for the
system $^{64}$Ni$+^{96}$Zr is compared with the data of
\citeasnoun{stef92}. Here the results of coupled-channels calculations
to all orders (solid lines) agree with the data very well as opposed
to the no-coupling (the dotted lines), linear-coupling (the dot-dashed
lines), and quadratic coupling (the dashed lines) cases. The upper
panel compare theory and calculations for the fusion cross section and
the lower panel for the average angular momenta. In this calculation 
up to double phonon states are included in the coupled channels. 
An important feature
of this calculation is that a truncation of the coupling within 
the double phonon space even at the
quadratic level is not sufficient to describe the data, one needs to
include couplings to all order.

\subsection{Angular Momentum Distributions}

\citeasnoun{diG91} also compared average angular momenta obtained
using different experimental techniques with theoretical
predictions. They found that there is good agreement between theory
and data obtained from isomer ratio and gamma-ray multiplicity
measurements with the exception of more symmetric systems, but not for
the fission fragment measurements. From the fission fragment angular
anisotropy measurements one obtains not $<\ell>$, but
$<\ell^2>$. Especially if the $\sigma_{\ell}$ distribution is pushed
to higher $\ell$ values as a result of coupling to other channels
$<\ell^2>^{1/2}$ may significantly differ from $<\ell>$, which may
explain some of the reported discrepancies. An excellent review of the
efforts to measure angular momentum distributions in fusion reactions
was given by \citeasnoun{vand1}.

There have been many attempts to extract average angular momenta
directly from the fusion excitation functions
\cite{reimer,reisdorf,neil2,serdar}. Moments of angular momenta are
related to the moments of fusion cross sections \cite{serdar}. These
relations can help assess the consistency of accurate fusion cross
section measurements with measurements extracting angular momentum
distributions using different methods.

\subsection{Probing Shape Phase Transitions with Fusion}

Until recently little attention was paid to subbarrier fusion on gamma
unstable targets. The Os and Pt region is interesting to study since
these nuclei go through a shape transition from prolate to oblate as
one increases the number of protons from 76 to 78.  $^{192}$Os has a
positive (prolate) quadrupole deformation parameter and a negative
hexadecapole deformation parameter. $^{194}$Pt has a quadrupole
deformation parameter similar in magnitude to those of $^{192}$Os, but
with a negative sign (oblate) and a hexadecapole deformation parameter
comparable to that of $^{192}$Os in sign and magnitude.  The isotopes
are similar in all respects other than the $\beta_2$ sign.  The effect
of this shape phase transition on the barrier distributions would be
noticeable by the skewness toward higher energies for prolate nuclei
and toward lower energies for oblate nuclei \cite{ibm4}. The barrier
distributions calculated using the IBM based model of Section \ref{IV}
by \citeasnoun{ibm4} is shown in Figure \ref{Fig6a}.

To understand this effect of the shape phase transition in the barrier
distributions, the fusion cross sections for transitional nuclei Pt
and Os were recently measured using an $^{16}$O beam by the Legnaro
group \cite{stef2} and a $^{40}$Ca beam by the Seattle group
\cite{prolateoblate}. (Since $^{40}$Ca is a heavier projectile one
expects this effect to be enhanced). The total fusion cross sections
calculated by \citeasnoun{ibm4} for the $^{16}$O$+^{194}$Pt system
agree very well with the evaporation residues measured by
\citeasnoun{stef2} once the fission cross sections estimated by the
statistical model calculations are subtracted.

The $^{40}$Ca$+^{194}$Pt and $^{40}$Ca$+^{192}$Os cross sections
measured and the associated barrier distributions extracted by
\citeasnoun{prolateoblate} are shown in Figure \ref{Fig6}, where the
results are also compared with the CCDEF calculations.  The
calculations take the excitations of the target nucleus into account
within the rotational model including both quadrupole and hexadecapole
deformations.  They also take into account the excitation of the
projectile to the 3$^{-}$ state at 3.7 MeV and the two neutron
transfer reactions from the target nucleus to the ground state of
$^{42}$Ca. The constant coupling approximations have been used for
vibrational excitation of the projectile and the transfer reactions,
while the radial dependence of the form factor of the collective model
was used for rotational excitations.  The predicted skewness of the
barrier distribution toward higher energies for the prolate nucleus
$^{194}$Os and toward lower energies for the oblate nucleus $^{194}$Pt
is observed. The second peak for both systems is due to the excitation
of the octupole state in the projectile. The barrier distributions for
both systems also exhibit a tail at the lower energies, which is not
reproduced by the CCDEF calculations. This problem was recently
associated with the contribution of the 2n transfer reactions to the
first excited 2$^{+}$ state of $^{42}$Ca \cite{prob1}.

One should point out that fusion barrier distribution extracted 
from the $^{16}$O$+^{186}$W data of \citeasnoun{lemmon} has the 
shape expected for a target nucleus with a negative hexadecapole 
deformation. There are pronounced differences between this 
distribution and that of $^{16}$O$+^{154}$Sm \cite{wei}, which 
are just those expected from a change in sign of the hexadecapole 
deformation. These data hence demonstrate s strong sensitivity of 
fusion to the hexadecapole deformation.  

\subsection{Difficulties in Extracting Barrier Distributions}

Since barrier distributions include the second energy derivative of
the cross section, very accurate measurements of excitation functions
at closely spaced energies are required. Even with very high precision
data, smooth barrier distributions can only be obtained under certain
model dependent assumptions such as fine-tuning the energy spacing for
calculating second derivatives \cite{izu,deweerd}. \citeasnoun{krr}
suggested to use integrals, rather than the derivatives of the fusion
data to improve model independence in the analyses. Unfortunately,
moments of the cross sections are even more featureless than the cross
sections themselves \cite{serdar}. In contrast, one of the main
advantages of using barrier distributions is that they bring out
important features in the data. However, moments of the cross 
section could be useful as they are related to the moments of 
angular momenta under certain assumptions
\cite{reimer,dasso3,esbensen2,serdar}.

Even with a very high precision in the fusion cross section it is
rather difficult to extract fusion barrier distributions at higher
energies as the cross section changes very slowly and the errors 
on the barrier distribution grow with energy. To illustrate the 
reason for this behavior consider a set of fusion data measured at 
a fixed energy spacing $\Delta E$. The second derivative may be 
approximated by the point-difference formula \cite{neil3} 
\begin{equation}
  \label{eq:pdif}
  \left( \frac{d^2(E\sigma)}{dE^2)} \right)_n = - 
\frac{2(E\sigma)_n - (E\sigma)_{n+1} - (E\sigma)_{n-1}}{(\Delta E)^2} .
\end{equation}
If the statistical errors on the cross section are a fixed fraction 
of their measured values
\begin{equation}
  (\delta \sigma)_n = f \sigma_n,
\end{equation}
then the error in the second derivative is
\begin{equation}
  \label{eq:sderr} 
  \delta \left( \frac{d^2(E\sigma)}{dE^2} \right) \sim 
\frac{\sqrt{6} f E \sigma}{(\Delta E)^2}.
\end{equation}
Hence the error on the second derivative increases as the cross 
section increases, whereas the second derivative itself gets smaller 
at higher energies.  Furthermore $\Delta E$ must be small enough to 
resolve any interesting structure, which also contributes to the 
large errors at high energies. 
  
It has been suggested that information about the barrier distributions
may be contained in the quasi-elastic scattering excitation functions
at backward angles \cite{kruppa,andres}. \citeasnoun{timmers} recently
developed a method to extract a representation of the fusion barrier
distribution from quasi-elastic excitation functions. They found that
although this representation of the quasi-elastic scattering data
indeed shows the general features of the fusion barrier distributions,
its sensitivity is reduced at high energies. More recently
\citeasnoun{elasticrowley} showed that the effects of strong coupling
are present in the barrier distributions from the elastic scattering,
but are smoothed out since different eigen-barriers have phase
differences.  Furthermore the effects are also smoothed by weak
couplings, which appear in first order in the elastic scattering cross
section, but only in second order in fusion cross section.  It would
be important to treat the phase problem properly in order to obtain
information on barrier distributions from the elastic scattering data.

\subsection{Fusion of Unstable Nuclei}

Heavy-ion fusion reactions induced by a halo nucleus or by an unstable
neutron rich nucleus are very intriguing current subjects of nuclear
physics \cite{ishcon}. Several groups have performed experiments to
examine whether the fusion cross section in such cases is
significantly different from that in heavy ion collisions induced by
the corresponding stable isotopes. \citeasnoun{yoshida} studied the
fusion reactions of $^{11,10,9}$Be with $^{209}$Bi at energies near
the Coulomb barrier. They observed no significant difference in the
excitation function for the collision of $^{11}$Be from that of
$^{10}$Be. On the other hand, \citeasnoun{fokou-y} have reported that
the induced fission cross section near the Coulomb barrier is much
larger in the $^{11}$Be + $^{238}$U reaction than that in the $^{9}$Be
+ $^{238}$U reaction. This is a very interesting result, though it is
not clear yet whether the fission took place via a compound nucleus
formation, suggesting an enhanced fusion cross section in the case of
unstable isotope.

\citeasnoun{taki91} and \citeasnoun{taki92} suggested that the fusion
cross section will be significantly enhanced if one uses a halo
nucleus as the projectile. In deriving this conclusion, the existence
of a stable soft dipole oscillation of the core nucleons against the
halo neutrons was assumed.  Though it is not yet completely settled
down, the existence of a physical soft dipole oscillation in light
halo nuclei is unlikely \cite{sagawa}. This might explain why the
fusion data of \citeasnoun{yoshida} do not show any characteristically
different behavior for the case of $^{11}$Be projectile.  As has been
shown by \citeasnoun{taki91}, the neutron halo itself can enhance the
fusion cross section by statically lowering the fusion barrier.  This
effect alone is, however, not so drastic.  Moreover, $^{11}$Be has
less pronounced halo property than $^{11}$Li.

Using time-dependent Hartree-Fock theory \citeasnoun{kim} showed that
nucleon transfer is enhanced for fusion reactions  between a stable
and an unstable nucleus with neutron halo. 

Some debate exists concerning the effects of break up of the halo
nuclei.  \citeasnoun{taki93} have shown that although the large
enhancement of the fusion cross section is moderated by the break-up,
the halo nucleus $^{11}$Li still leads to a larger fusion cross
section than the other Li isotopes. One should point out that this
conclusion relies on the assumption that there exists a soft dipole
resonance in $^{11}$Li.  On the other hand, \citeasnoun{canto} argued
that the fusion cross section of a halo nucleus will be hindered by
the break-up effect. To the contrary, \citeasnoun{dasso4} contended
that the break-up channel enhance the fusion probability.  In order to
reach a definite conclusion, one needs to know more about the radial
dependence of the break-up form factor and has to treat both the real
and the virtual break-up precesses in a consistent way including the
associated potential renormalization.

One interesting subject to be explored is the effects of the bond
formation due to halo neutrons on the fusion cross
section. \citeasnoun{carlos} studied this problem by using the fusion
between $^{11}$Li, consisting of $^{9}$Li core and halo di-neutrons,
and $^{9}$Li as an example .  As $^{11}$Li and $^{9}$Li approach each
other, there is a particular separation distance at which both the
Coulomb and nuclear potentials between the two cores are small, but
the two neutrons in the halo can be shared by both cores. One can then
investigate the effect of this molecular bonding on the fusion of
$^{11}$Li and $^{9}$Li. These preliminary calculations indicated
\cite{carlos} a very significant enhancement of the fusion cross
section due to molecular bonding. \citeasnoun{padova} showed that this
cross section is somewhat reduced when different initial conditions
are used. The calculations of both groups were done in the adiabatic
approximation, which tends to overestimate the effect.  Existence of
molecular bonding, and its effect on the fusion process remains to be
an open question.

In this connection, we wish to remark that a polarization of the wave
functions of the valence neutrons, i.e., the admixture of higher
orbits, is essential in order for the bond effect to be
significant. \citeasnoun{imanishi} studied the bond effect in the
fusion between $^{11}$Be and $^{10}$Be and showed that a significant
polarization of the valence nucleons starts to take place even where
the core nuclei are still far apart if the binding energy of the
valence nucleons is small, and that consequently there exists a large
bond effect in this system. Notice that $^{11}$Be has two bound states
{1p$_{1/2}^{-}$} and {2s$_{1/2}^{+}$} and one low lying resonance
state {1d$_{5/2}^{+}$}.  The hybridization of these opposite parity
configurations causes a large polarization.

It is possible that the response of the projectile in heavy ion
collisions induced by a neutron rich unstable nucleus can be
formulated in terms of the coupling to a resonance state.  Several
models exist to calculate the effect of the width of the resonant
state on quantum tunneling \cite{ap,hus95}.  

\subsection{Fusion of Very Massive Systems and Superheavy Nuclei}

As the compound nuclei formed in the fusion get heavier, fission
becomes an increasingly important de-excitation channel. For such
reactions, evaporation residues and fission fragments must be added to
obtain the total fusion cross section. As systems get more massive
($Z_pZ_t > 1000$), fusion starts competing with other reaction
channels representing a significant exchange of energy, charge, mass,
and angular momenta. Understanding the dynamics of fusion and
competing reactions for very massive systems is essential, among other
things, to assess the conditions for the formation of super-heavy
elements. These topics are covered in a recent review by
\citeasnoun{reisdorfrev}.

A significant difference between the fusion of massive nuclei and the
fusion of medium weight nuclei which we discussed so far is that the
fusion cross section for very massive systems is not enhanced, but
rather hindered. This situation is encountered when the product of the
atomic numbers of the projectile and target exceeds about 1800
\cite{reisdorfrev}.  The incident energy has to be considerably higher
than the fusion barrier expected from the Bass potential \cite{bass}
which was determined to fit the fusion data above the barrier for
medium weight systems with $Z_pZ_t = 64-850$ in a potential
model. This excess energy is called the extra push energy.
\citeasnoun{bjornholm} attributed this hindrance phenomena to the fact
that the fission barrier for massive systems locates well inside the
potential barrier in the entrance channel, and introduced the concepts
of the extra push and the extra extra push. The former is the energy
needed to overcome the conditional saddle, i.e., the saddle under the
constraint of mass asymmetry, while the latter is the energy needed to
carry the system inside the unconditional saddle for fission. Though
there have been quite a number of experimental as well as theoretical
studies of the extra push energy, its origin and the dependence on
various parameters of the system, such as the effective fissility
parameter in the entrance channel, are not fully understood.  One
should note that the decrease of the fusion cross section with
decreasing bombarding energy in massive systems, where there exists an
extra push, is also much slower than that expected in the potential
model where there exists an extra push. This indicates the existence
of a kind of enhancement mechanism of the fusion cross section in
massive systems as well once the hindrance effect associated with the
extra push problem is isolated.

An interesting problem concerning the fusion of massive nuclei is the
synthesis of super heavy elements.  A significant advance occurred
when \citeasnoun{gsi} at GSI working with the SHIP velocity filter has
succeeded in producing the super heavy element Z=112 by the so called
cold fusion method using $^{70}$Zn + $^{208}$Pb reaction. This is the
heaviest element recorded to date and it is only two atomic number
away from the predicted magic number Z=114. Though the successful
synthesis of element 112 after the synthesis of elements 110 and 111
in 1994 seems to indicate that a similar experimental strategy can be
used to go further to the realm of the heaviest elements, a problem is
that the cross section is very small, i.e., of the order of 1
pb. Actually, only two events were identified for $Z=112$.  It would
certainly be very interesting to look for alternative ways to
synthesize super heavy elements. An interesting question in this
connection is to examine whether there are advantages of using neutron
rich unstable nuclei. A preliminary study in this direction has been
undertaken by \citeasnoun{taki92} and \citeasnoun{taki92c}. These
authors discussed the advantages such as the larger survival
probability of the compound system, lowering of the fusion barrier,
and the possible lowering of the extra-push energies, and
disadvantages such as the low beam intensity in reactions induced by
neutron rich unstable nuclei.  In passing, we wish to mention that
\citeasnoun{nomura} is trying to use (HI, $\alpha$xn) reactions to
experimentally synthesize super heavy elements using the cooling
mechanism by $\alpha$ particle emission, and that \citeasnoun{aritomo}
are introducing a diffusion model to theoretically discuss the
mechanism of the synthesis of super heavy elements, though both of
them treat a thermal process rather than a quantum tunneling process.

\section{Open Problems and Outlook}
\label{VI}

Although there still are many unsettled issues even in fusion
reactions with stable nuclei, remarkable progress has been made in the
last fifteen years. New, conceptually alluring ways of analyzing data,
such as studying barrier distributions; new approaches to channel
coupling, such as the path integral and Green function formalisms; and
alternative methods to describe nuclear structure effects, such as
those using the Interacting Boson Model, were introduced. The roles of
nucleon transfer, higher-order couplings, and shape-phase transitions
were elucidated. We can now understand the data for clean (i.e.,
asymmetric) systems in terms of inelastic excitations and nucleon
transfer.  Acquisition of high precision, complementary data for
fusion, transfer reactions, and elastic scattering at below and near
the barrier should be encouraged as theoretical tools are available to
analyze them.

On the other hand, fusion cross sections for very heavy symmetric
systems cannot be reproduced by the present models. In such systems
inclusion of higher-order coupling is essential.  One salient
ingredient is a proper description of neck formation. Though there are
many pioneering attempts which relate the large enhancement of the
fusion cross section to the neck formation
\cite{jahnke,krappe,iwamoto,iwamoto2,aguiar}, a microscopic
description of fusion reactions in general and neck formation in
particular is still at a very primitive stage and needs to be further
developed. In connection with the former, we wish to note the computer
simulations for sub-barrier fusion reactions by
\citeasnoun{bonasera}. The effects of neck formation could be
formulated in terms of the quantum tunneling in a multi-dimensional
space \cite{nix,kodama,denisov1,denisov2}.

Several existing and presently under construction experimental
facilities providing beams of short-lived radioactive nuclei present
new opportunities to explore the dynamics of the fusion reactions
below the Coulomb barrier. In such facilities, in addition to testing
our present understanding of the fusion dynamics in a new setting, one
can experimentally investigate entirely new facets such as the
coupling of resonant states to quantum tunneling and the possibility
of molecular bond formation.

One should finally remark that there are many other tunneling
phenomena in nuclear physics besides heavy-ion fusion reactions. Alpha
decay, fission, various rare decays, and nuclear structure problems
such as the decay of a superdeformed band to a normal band will also
be effected by coupling to the intrinsic degrees of freedom and
insight obtained in the study of heavy-ion fusion reactions at
subbarrier energies will be a valuable tool to understand these
phenomena in more detail.

\acknowledgments

We thank J. Beacom and K. Hagino for their comments on the manuscript.
We are grateful to our collaborators and colleagues J. Bennett,
A. DeWeerd, K. Hagino, S. Kuyucak, J. Leigh, N. Rowley, R. Vandenbosch
for many discussions over the years. This research was supported in
part by the U.S. National Science Foundation Grants No. PHY-9314131
and PHY-9605140; in part by the Japan Society of Promotion of Science;
and in part by the University of Wisconsin Research Committee with
funds granted by the Wisconsin Alumni Research Foundation. It was also
supported in part by the Grant-in-Aid for General Scientific Research,
Contracts No. 06640368 and No. 08640380; the Grant-in-Aid for
Scientific Research on Priority Areas, Contracts No. 05243102 and
No. 08240204; and Monbusho International Scientific Research Program:
Joint Research from the Japanese Ministry of Education, Science and
Culture Contract No. 09044051.

\newpage

\begin{figure}[t]
\caption{One dimensional potential of Eq. (4) for the
$^{64}$Ni$+^{64}$Ni system for several $\ell$ values. The lowest
barrier is for $\ell=0$ (the bare barrier). The middle and top
barriers are for $\ell=100$ and $\ell=150$ respectively.            }
\label{Fig0}
\end{figure}

\begin{figure}[t]
\caption{The effective radius extracted from fusion calculations for
the $^{16}$O+$^{154}$Sm system using Eq.~(6). The curves
correspond to spherical, vibrational and deformed nuclei with quadrupole
coupling strengths $v_2=0$, 0.13 and 0.26, respectively. From
Balantekin {\em et al.} (1996).} 
\label{Fig1}
\end{figure}

\begin{figure}[t]
\caption{Classical (on the left) and quantum mechanical (on the right)
transmission probabilities for a one-dimensional potential barrier.}
\label{Fig2a}
\end{figure}

\begin{figure}[t]
\caption{Classical (on the left) and quantum mechanical (on the right)
transmission probabilities for a two-channel coupling. $V_0$ is the
height of the one-dimensional potential barrier coupled to these
channels.}
\label{Fig2b}
\end{figure}

\begin{figure}[t]
\caption{Fusion cross section and barrier distribution for the 
$^{16}$O + $^{154}$Sm system by \protect\citeasnoun{lei95}.}
\label{Fig3}
\end{figure}

\begin{figure}[t]
\caption{Effective one-dimensional potential barriers from (Balantekin
{\em et al.}, 1983). The outer turning points are determined from the
phenomenological potential of Krappe {\em et al.} (1979) to fit the
peak positions. The distance between the outer and inner turning
points is the thickness function inverted from the data. The shaded
region indicates the error envelope. The short dashed line denotes the
point Coulomb potential and the long-dashed line denoted the potential
of (Krappe {\em et al.}, 1979).}
\label{Fig4}
\end{figure}

\begin{figure}[t]
\caption{Illustration of the geometric interpretation of the sudden
approximation. The solid curve is the total potential for the
$^{16}$O$+^{154}$Sm system when the projectile is taken to be
spherical. The dashed ($\lambda=-0.327$) and dot-dashed
($\lambda=+0.613$) curves are potentials for the two-level
approximation for the target with $\beta=0.25$. The arrows show the
shifts predicted for the barrier peaks \protect\cite{serdar}.}
\label{Fig4a}
\end{figure}

\begin{figure}[t]
\caption{A systematic study of subbarrier fusion of $^{16}$O
projectile with rare earth nuclei, the structure of which is described
using the Interacting Boson Model (see text).}
\label{Fig5}
\end{figure}

\begin{figure}[t]
\caption{Fusion cross sections from \protect\citeasnoun{ack96} for the
systems $^{58}$Ni$+^{64}$Ni (circles) and $^{64}$Ni$+^{64}$Ni
(squares) as a function of the energy normalized to the barrier
height.}
\label{Fig6b}
\end{figure}

\begin{figure}[t]
\caption{Comparison between measured and calculated barrier
distributions for $^{36}$S$+^{110}$Pd system as more phonons are
included in the calculation. The data are from
\protect\citeasnoun{stef4} and the calculation is from
\protect\citeasnoun{deweerd}. The dot-dashed, dashed, and solid lines
correspond to calculations including one-phonon, two-phonon, and
three-phonon states, respectively.}
\label{Fig6c}
\end{figure}

\begin{figure}[t]
\caption{Fusion cross section (upper panel) and the average angular
momenta (lower panel) for the $^{64}$Ni$+^{96}$Zr system. The data is
from \protect\citeasnoun{stef92}. the theoretical calculation is from
\protect\citeasnoun{hag97}. The two-phonon states of the quadrupole
surface vibration of both the projectile and the target are taken
into account in the coupled channels calculations. The dotted line is
the result in the absence of channel coupling. The dot-dashed and
dashed lines are the results when the nuclear potential is expanded up
to the first and the second order terms in the deformation parameters,
respectively. The solid line is the result of the coupled channels
calculations to all orders, obtained without expanding the nuclear
potential.}
\label{Fig7}
\end{figure}

\begin{figure}[t]
\caption{Predicted behavior of the barrier distributions for fusion
reactions on the prolate ($^{194}$Os) and the oblate ($^{194}$Pt)
nuclei \protect\cite{ibm4}.}
\label{Fig6a}
\end{figure}

\begin{figure}[t]
\caption{Experimentally determined fusion cross sections by
 \protect\citeasnoun{prolateoblate} for prolate and oblate nuclei. The
 solid curve is the simplified coupled channels calculation with the
 CCDEF code. The dashed curve is the result for a one-dimensional
 barrier ignoring all the couplings.}
\label{Fig6}
\end{figure}

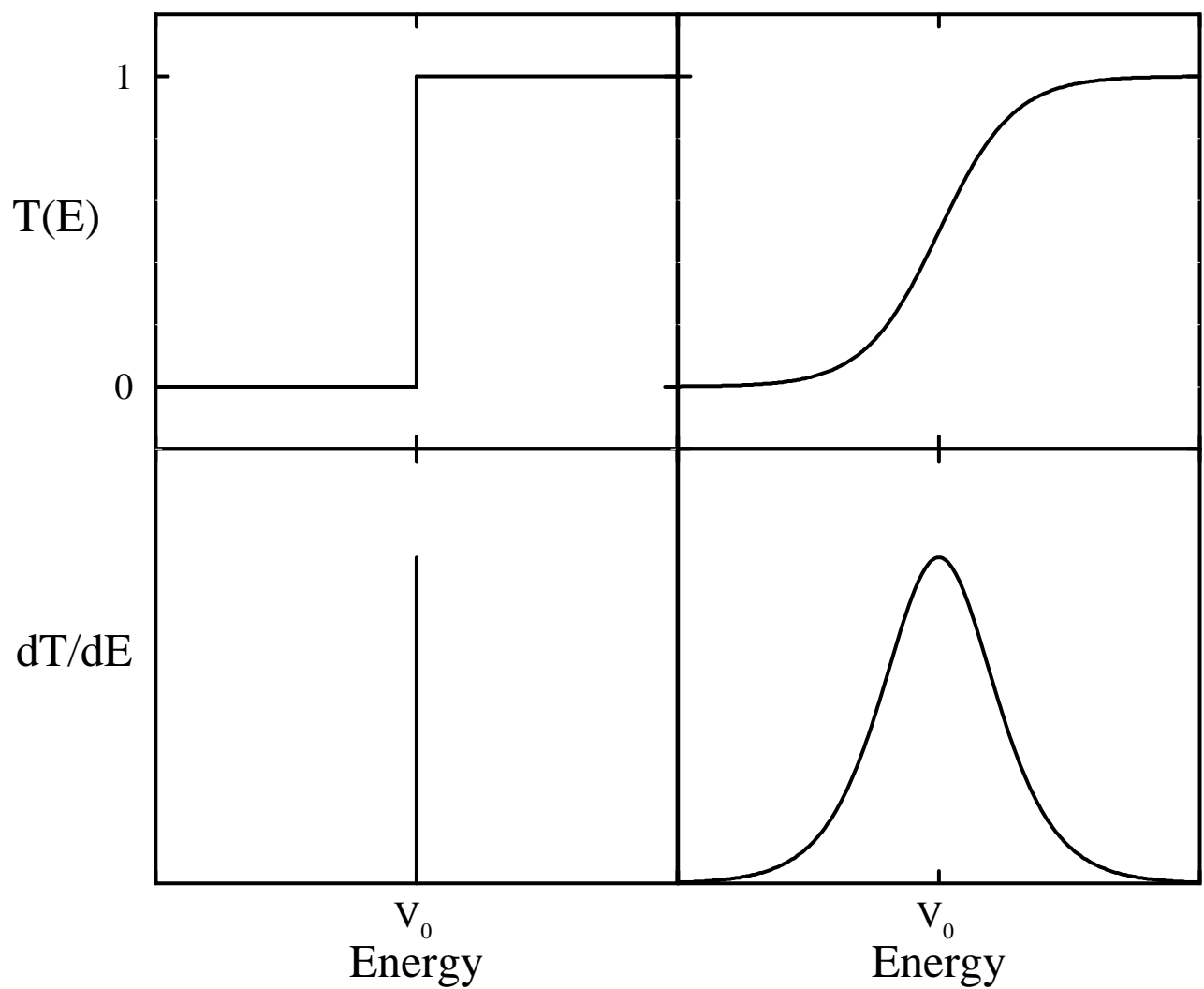
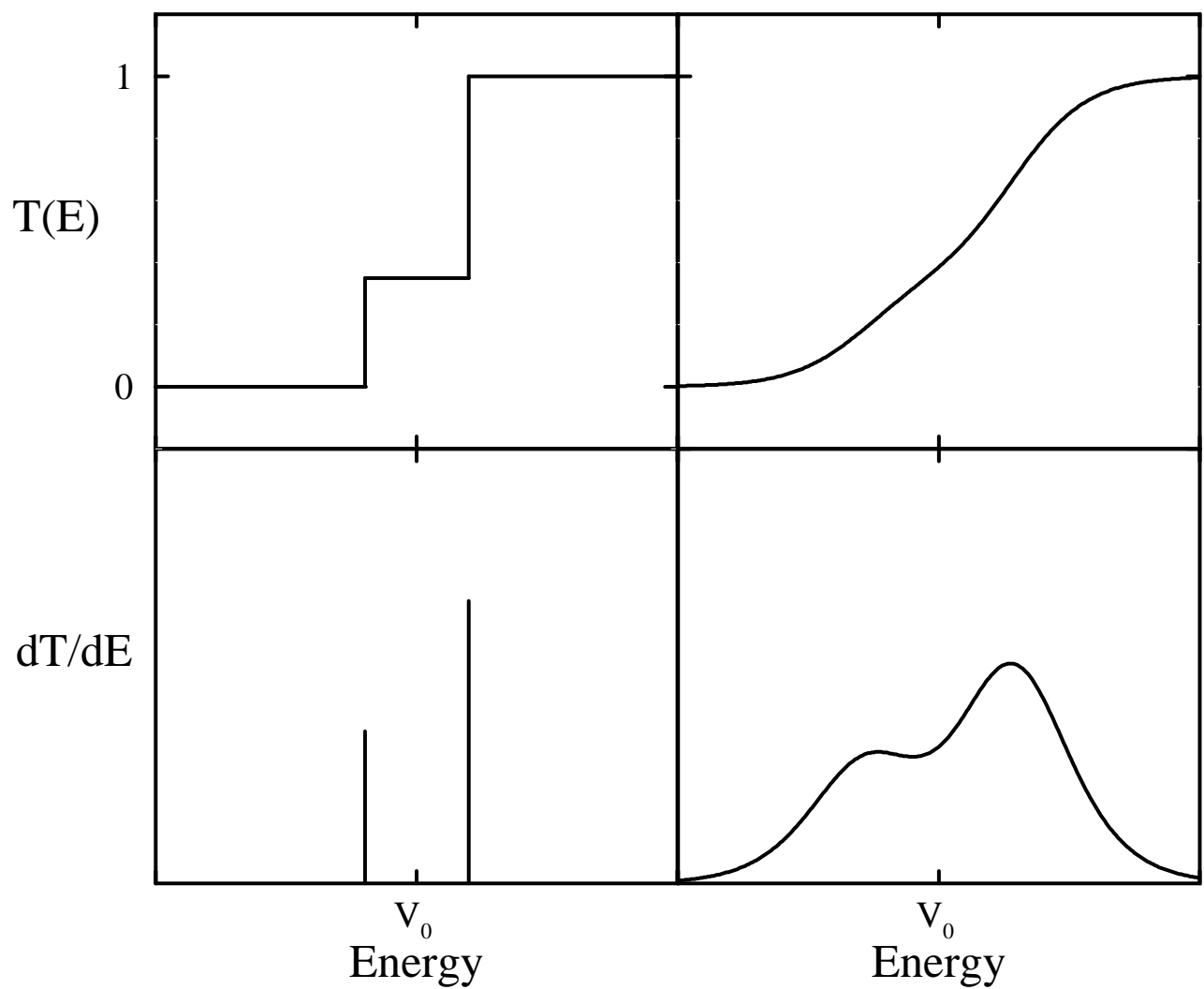
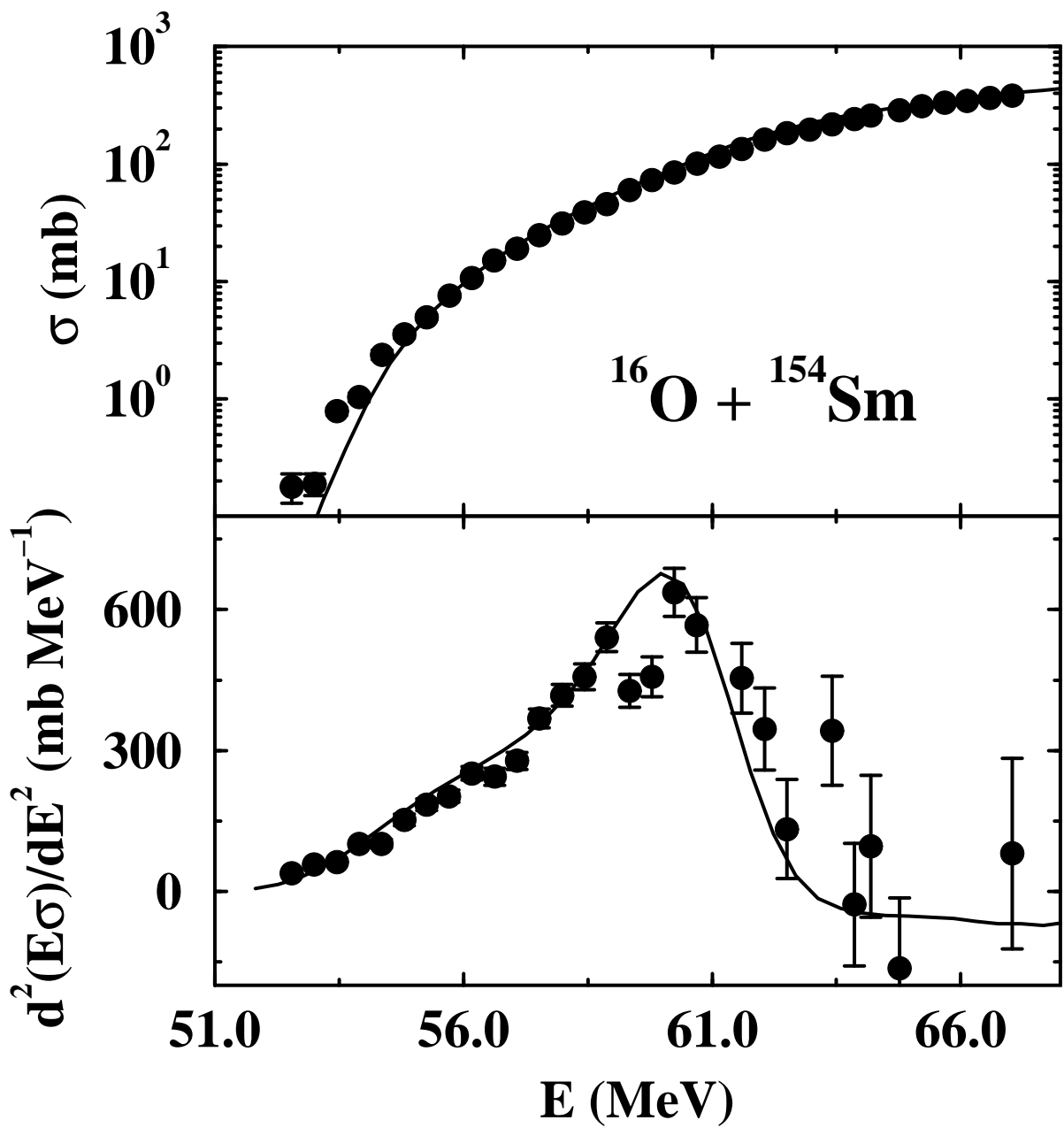
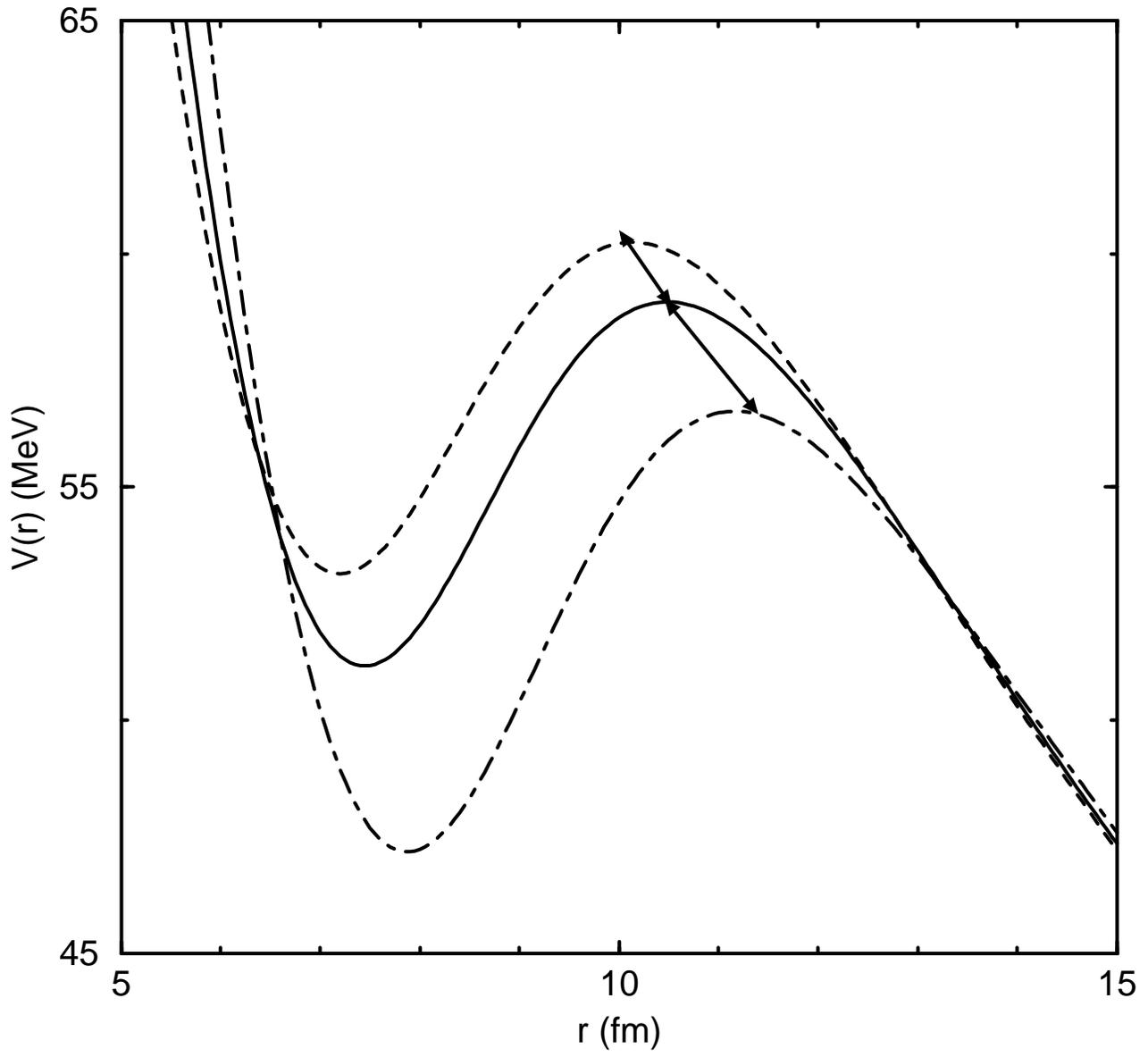
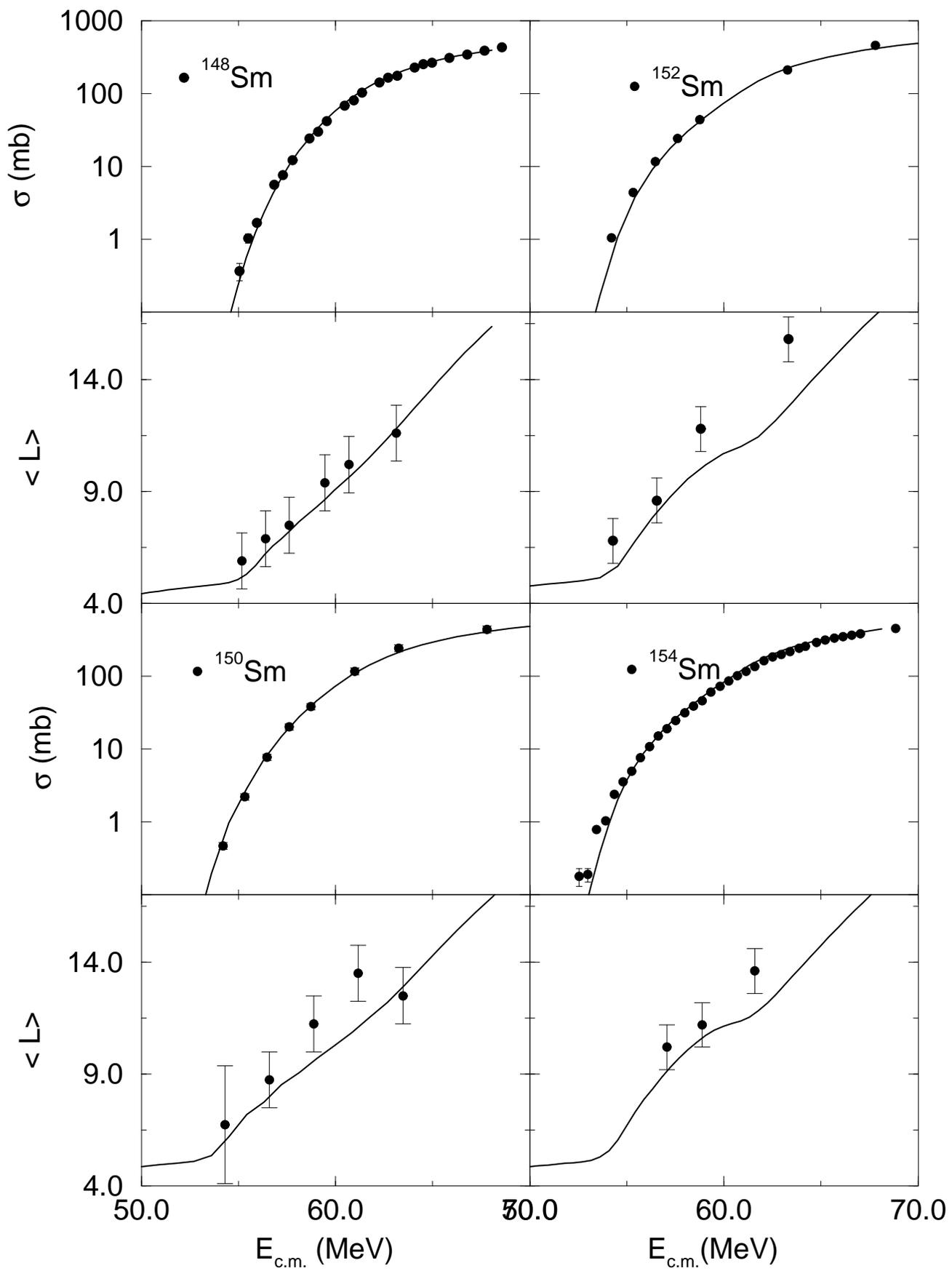
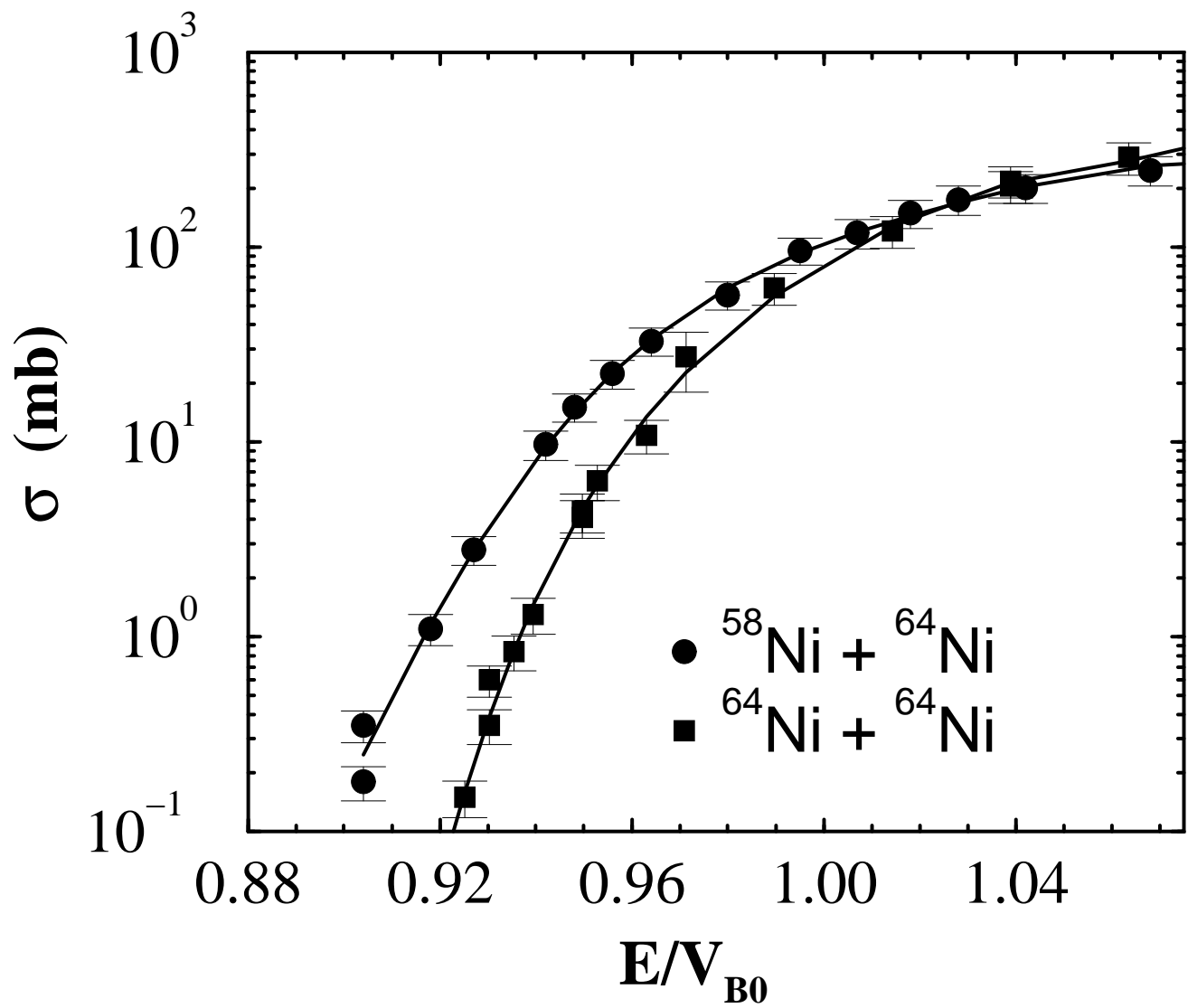
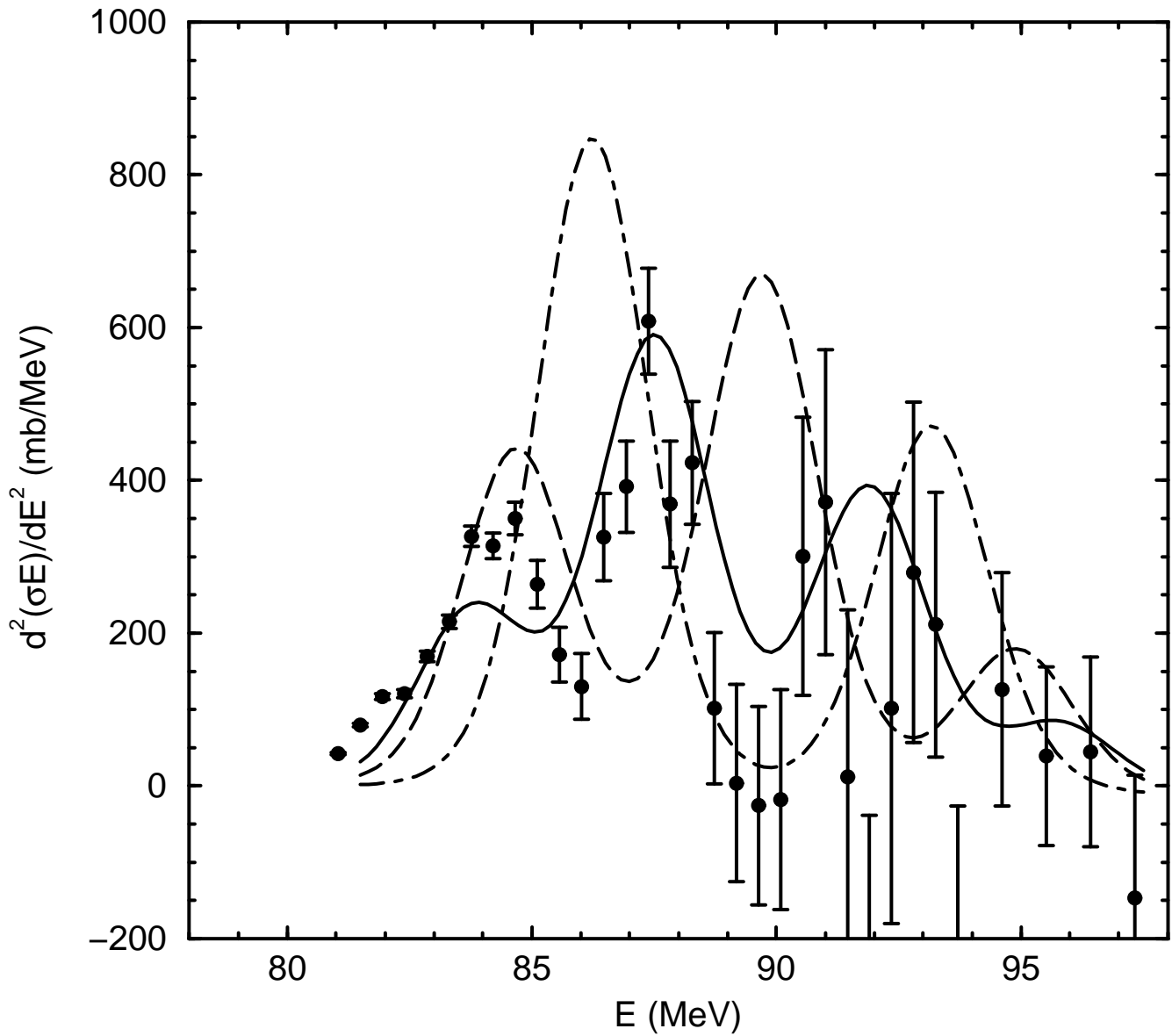
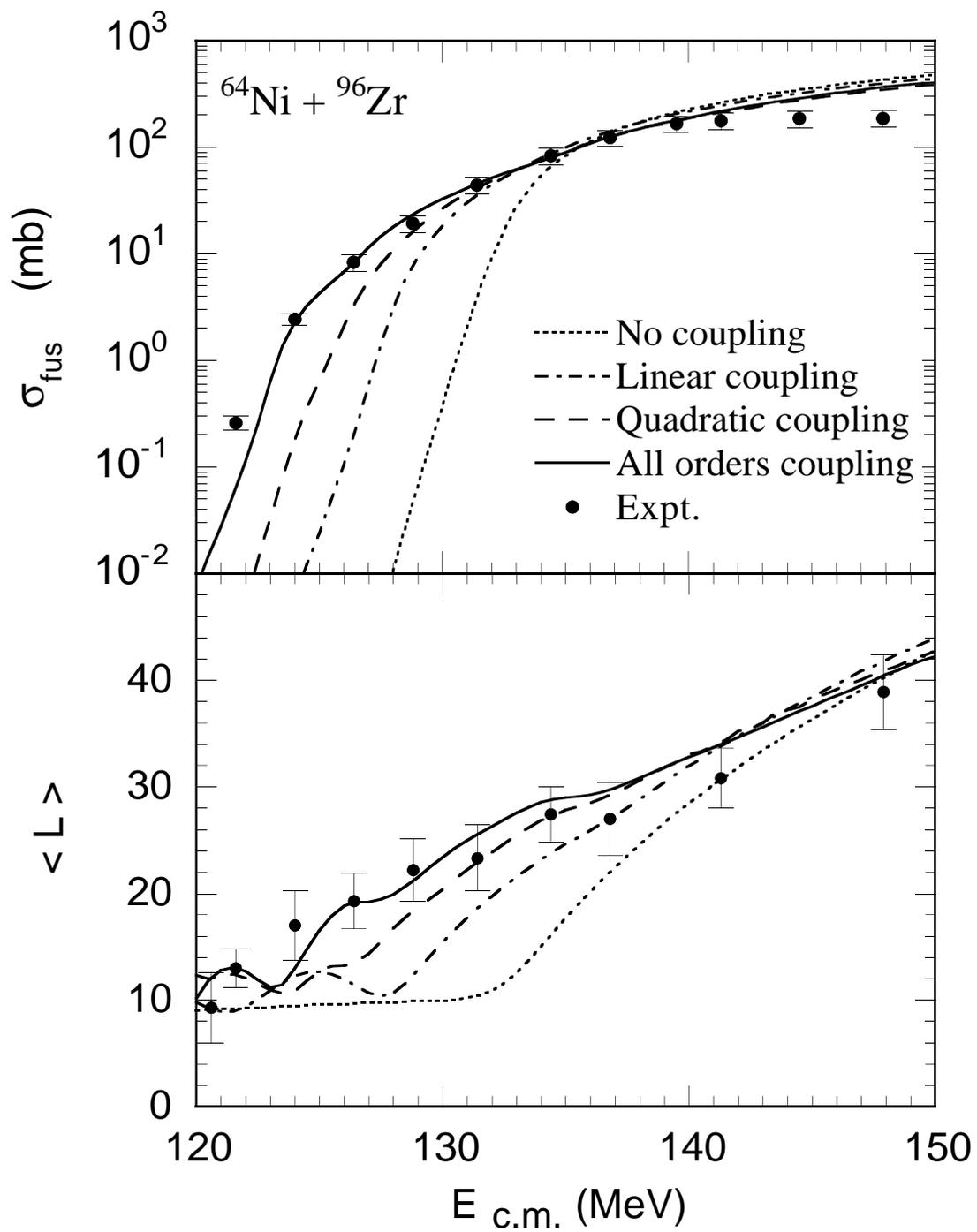
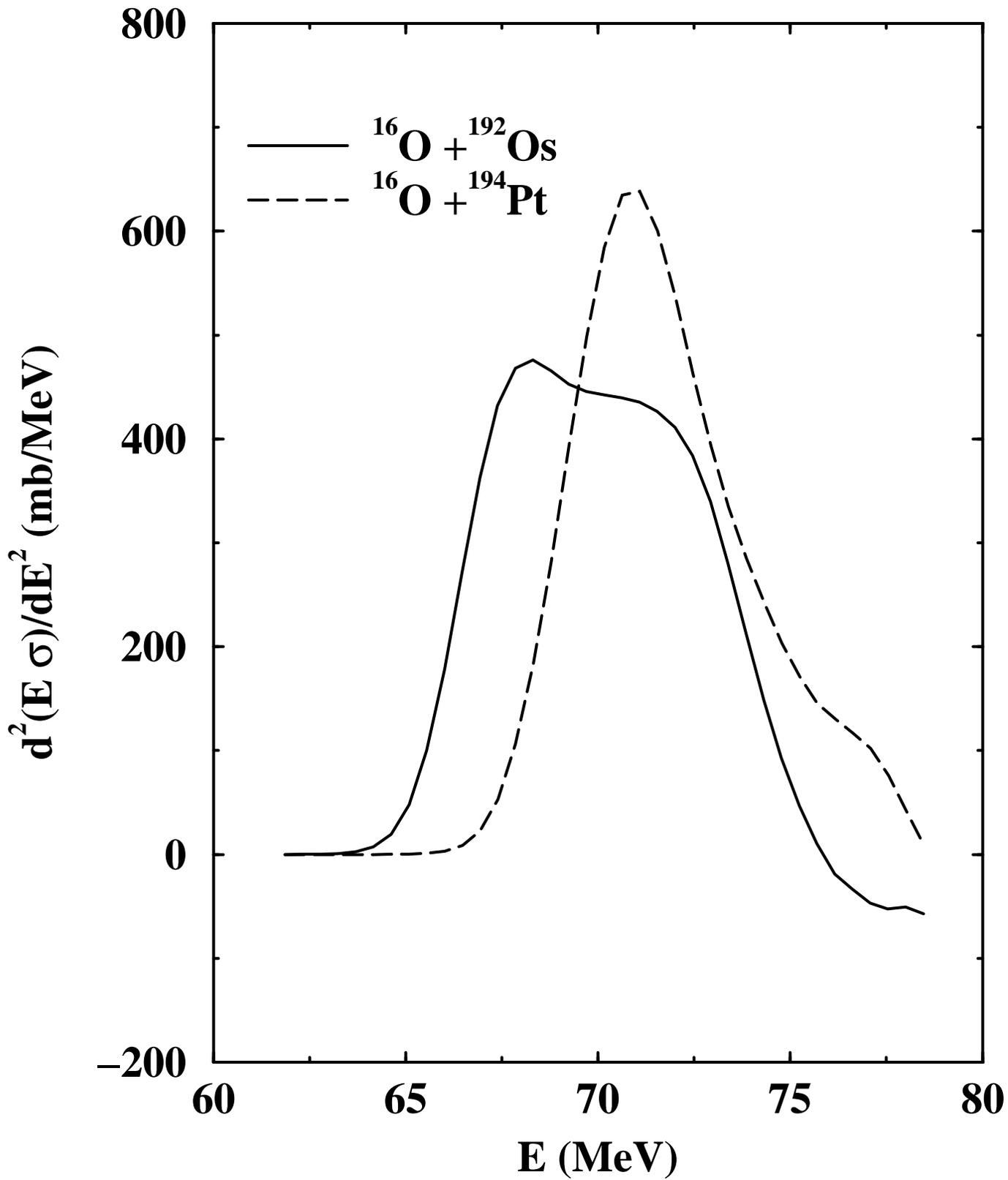
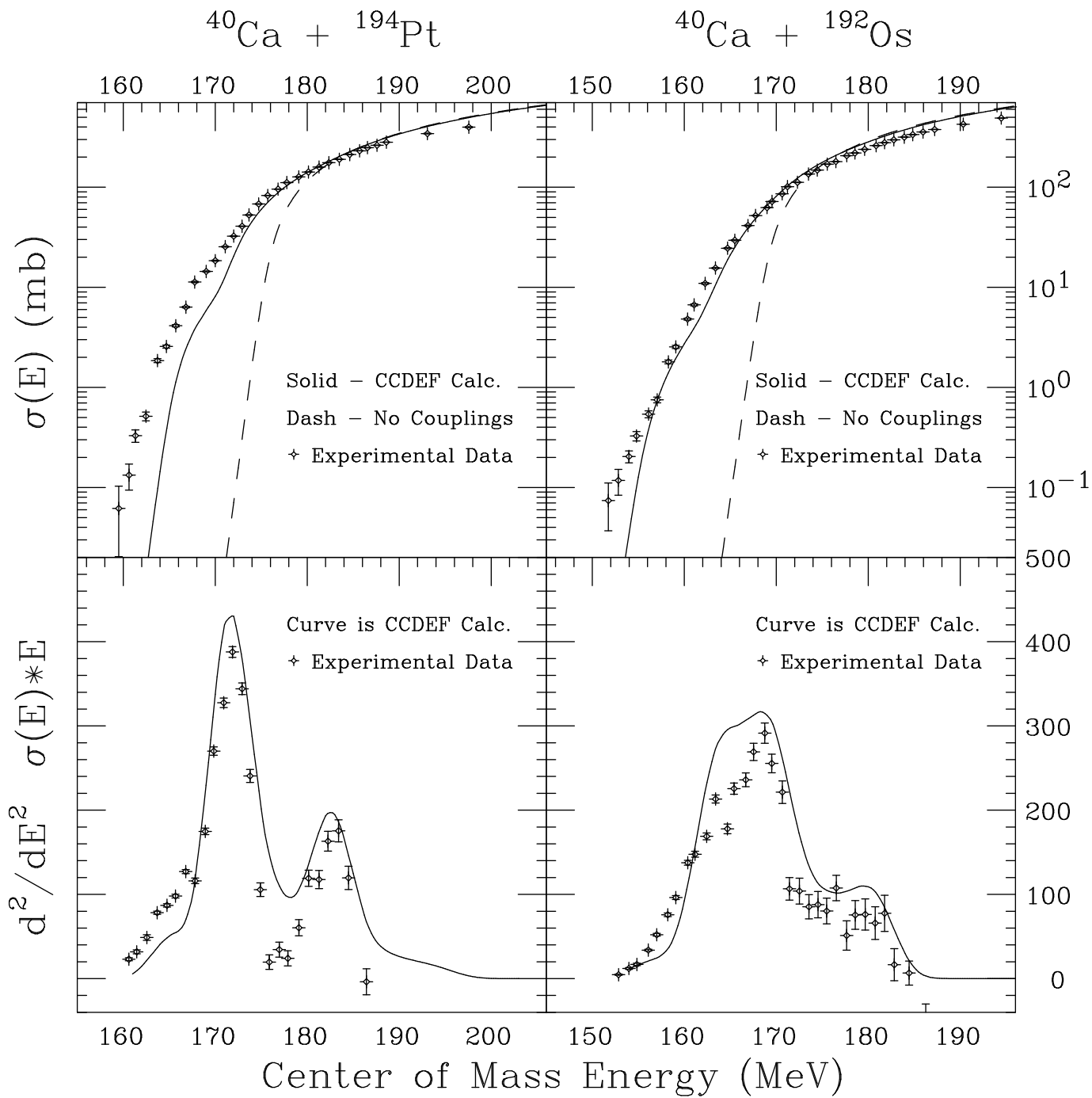
\end{document}